\documentclass[12pt]{article}             %Arxiv

\makeatletter %%REMOVE FOR SIAP
\renewcommand{\section}{\@startsection
{section} {1} {0mm} {-\baselineskip} {0.5\baselineskip}
{\large\bf}}
\renewcommand{\subsection}{\@startsection
{subsection} {2} {0mm} {-\baselineskip} {0.5\baselineskip}
{\normalsize\bf}} \makeatother

\usepackage{amsmath}
\usepackage{amssymb}
\usepackage{amsfonts}
\usepackage{amsthm}
\usepackage{latexsym}
\usepackage{epsfig}
\usepackage{array}
\usepackage{boxedminipage}

\newcommand{\vs}{\vspace{1mm}}
\newcommand{\vv}{\vspace{2mm}}

\newcommand{\vvvv}{\vspace{4mm}}

\newcommand{\K}{\mathbb{K}}
\newcommand{\C}{\mathbb{C}}

\newcommand{\R}{\mathbb{R}}

\newtheorem{defin}{Definition}
\newtheorem{theor}{Theorem}

\newtheorem{propo}{Proposition}

\newcommand{\ke}{\mathrm{ker} \hspace{0.5mm}}
\newcommand{\im}{\mathrm{im} \hspace{0.5mm}}

\newcommand{\dsp}{\displaystyle}

\newcommand{\tra}{{\sf T}}

\begin{document}

\begin{center}
%SIAM
%\title{%Linear 
%Nonlinear Stuff}
{\Large Associate submersions 
%fully-implicit characteristics in\vs\\ 
%nonlinear circuit theory}\vs\\
%fully implicit descriptions
and %bifurcations 
qualitative properties of nonlinear circuits\vspace{2mm}\\
with %fully-
implicit characteristics}\footnote{This is the author's final version of a paper accepted
for publication in the {\em International Journal of Bifurcation and Chaos}, 2019.} \vvvv \\
%Solvability of Nonlinear Circuits}\vvvv\\
  %Circuit Theory in Projective Space and \\
%Homogeneous Circuit Models}
%\author{
Ricardo Riaza\footnote{Depto.\ de Matem\'{a}tica Aplicada a las TIC
\& Information Processing and Telecommunications Center, 
ETS Ingenieros de Telecomunicaci\'{o}n, Universidad Polit\'{e}cnica de Madrid, Spain. {\sl ricardo.riaza@upm.es}.}\vv
%Research supported by Project MTM2015-67396-P (MINECO/FEDER).}}
%}
%END SIAM
\end{center}

%\date{} %\today}

%\markboth{IEEE Transactions on Circuits and Systems -- I REGULAR PAPERS
%(SUBMITTED)}
%~Vol.~X, No.~Y, Month-Year}%

%\maketitle

%\ifthenelse{\boolean{ieee}}{\documentclass[journal,twoside]{IEEEtran}}{\documentclass[12pt]{article}}

%\ifthenelse{\boolean{ieee}}{}{\mbox{}\vspace{-13mm}} %only arxiv

\begin{abstract}
%\vs

We introduce in this paper an equivalence notion
for submersions $U \to \R$, %$U \subseteq \R^2$,
%with 
$U$ open in $\R^2$, 
which 
makes it possible to identify 
%identifies 
a smooth planar curve with a unique class of 
submersions. This
idea, which extends to the nonlinear setting
the construction of a dual projective space,
provides a systematic %and general 
way to handle global implicit
descriptions of smooth planar curves.
We then 
apply this framework to model 
%(characteristic curves of) 
nonlinear electrical devices as {\em classes of equivalent functions}. 
%and 
%tackle with 
%use the resulting framework %formalism to 
%tackle certain analytical problems in nonlinear circuit theory.
%involving nonlinear circuits
%within the resulting framework. % framework so obtained.
In this setting, linearization naturally accommodates incremental 
resistances (and other
analogous notions) in homogeneous terms.
This approach, combined with a projectively-weighted version
of the matrix-tree theorem, makes it possible to formulate and address
in great generality several problems in nonlinear circuit theory.
In particular, we tackle unique solvability problems in resistive circuits, and discuss a 
general expression for the characteristic polynomial of dynamic circuits at equilibria. Previously known results, which were derived in the literature
under unnecessarily restrictive working assumptions, are simply obtained
here by using dehomogenization. Our results are shown to apply also to circuits
with memristors.
We finally present a detailed, graph-theoretic
study of certain stationary bifurcations in nonlinear circuits using the formalism
here introduced.

\end{abstract}

%%%%%%%%%[PARA EL ABSTRACT O INTRODUCTION: 
%In topologically nondegenerate problems,
%the homogeneous formalism paves the way for a completely general
%characterization of the so-called regular set in graph-theoretic terms
%(specifically, in terms of the family of spanning trees in the circuit)
%and, subsequently, of the regular manifold
%where the circuit equations define a smooth flow.

%\

%%%%%%%%%%%%[ABSTRACT: 
%nice distinction between linear and (in a strict and local sense) nonlinear circuits
%with regard to the structure of the so-called regular set: empty or the whole space in linear
%cases, open dense in locally nonlinear cases. And the latter are generic]

\vspace{2mm}

\noindent {\bf Keywords:} planar curve, projective space, homogeneous 
coordinates, matrix-tree theorem, nonlinear circuit, memristor, 
characteristic polynomial, equilibrium
point, bifurcation.

%\newpage
\section{Introduction}
\label{sec-intr}

%PDTE UNIFY j k PRODUCTS (THE SAME OR SPLIT \PROD)

Implicit descriptions are often used at the initial stages in dynamical
system modelling within many branches of science and engineering. Incidentally, 
implicit systems of differential-equations, also termed differential-algebraic systems or singular systems,
have become a widely used tool in system modelling,
with different names depending on the application field: 
e.g.\ DAEs are termed {\em semistate systems} in circuit modelling,
 {\em descriptor systems} in control theory, or {\em constrained systems}
in mechanics
\cite{bre1, kmbook, lmtbook, rabrheinmech, wsbook}.
Much research has been focused on numerical methods directed to such implicit 
descriptions: see \cite{bre1, kmbook, lmtbook, rabrheintheo}.

However, from an analytical perspective, it is
common to renounce %/give up 
such implicit formulations at a relatively early stage
of the aforementioned modelling process, in particular in the formulation of reduced-order
models.
This is the case when using local coordinates for the description of the manifold
where trajectories lie, within the aforementioned
differential-algebraic framework.
%(often, by means of the implicit function theorem or the subimmersion theorem)
%Other common scenarios involve the restriction
Alternatively, one can restrict the scope of the analysis to cases in which such a manifold
admits a global coordinate description. The price 
we pay for this is of course a loss of generality.
% a manifold... (IFT, solution manifold
%of a DAE) or restricting the
%attention to problems in which a global description in terms of a classical variable is
%assumed to exist. 

The circuit context provides clear examples of this. Focusing for instance on
a nonlinear
resistor (a device defined by a nonlinear characteristic relating voltage and current), 
it is very common to assume that this characteristic either has a global 
current-controlled form
or a global voltage-controlled form, i.e.\ that it can be written either as $v=h(i)$ or 
as $i=g(v)$ for appropriate, globally defined 
functions $h$ or $g$. Analogous remarks apply for other types of devices.
%capacitors, inductors and memristors.
%Memristive context: almost from the very beginning a flux-controlled form or charge-controlled form...
%\cite{ishaq2017}
%Characteristic polynomial in Ishaq paper (have PDF)
In many cases such restrictions make perfect 
sense because of physical reasons but, 
at the risk of being too sketchy, 
from a mathematical point of view they may also reflect the fact that
the available concepts and tools are mainly directed to such restricted
contexts. Keeping a fully-implicit form (that is, working with a general 
implicit characteristic $f(i, v)=0$ throughout the analysis, 
to continue with the example
above) often involves additional difficulties or is simply unfeasible.
The reader is referred to subsection \ref{discussion} %\ref{sec-bac} 
for a more detailed discussion of this.
Regardless of the reasons, it is clear that such a loss of generality 
is important from a theoretical point of view and
may also be relevant in many practical contexts,
in particular in qualitative studies of nonlinear systems (including e.g.\ global
bifurcation analyses).

Surprisingly, such unnecessarily restrictive assumptions often occur in the simplest
linear problems. 
%To continue with the circuit example, a linear resistor already exemplifies this; when 
In the circuit context, when
writing Ohm's law for a linear resistor 
as $v=zi$ or $i=yv$ (we use the common notations $z$, $y$ for the impedance
and the admittance parameters, which take complex values; 
the reader may also think of resistance and conductance in the real domain) 
some cases are necessarily left out; 
indeed, the form $v=zi$ does not accommodate an open-circuit,
whereas $i=yv$ excludes a short-circuit. The idea, of course, 
is to treat $z$ or $y$ as
symbolic variables in general parametric analyses, not
focusing on specific numerical values.
To this aim, we might consider using a fully-implicit description of a linear 
resistor or an impedance, by writing such a characteristic in the general
form 
\begin{equation} \label{homogOhm}
pv-qi=0, 
\end{equation}
with the only requirement that the parameters $p$ and $q$ do not vanish
simultaneously. %Sources rhs*** 
As detailed in subsection \ref{subsec-lindev}, it is natural to think
of $(p:q)$ as homogeneous coordinates of a point in a (real or complex) projective line, with the invariant ratios $y=p/q$ and $z=q/p$ defined on 
the corresponding subsets of the parameter space.
The key idea is to look at the left-hand side of (\ref{homogOhm}), 
with fixed values of $p$ and
$q$, only as a representative of a 
family of equivalent forms, which are defined up to a non-vanishing constant (as in the
standard construction of the projective space from a given vector space, in this case
the space of linear forms $L(\K^2, \K)$, with $\K=\R$ or $\C$). This way we 
{\em identify
a linear resistor with an equivalence class of linear forms} and we are naturally led
to a projective context. 
This supports a projective-based formalism to set up linear circuit models in broadly general 
terms, as detailed in \cite{homoglin}. % and summarized in Section \ref{sec-bac} **.

In this context, the present paper is driven by a two-fold goal.
%The purpose of the present Letter is 
Our first purpose is to set the foundations %to extend 
for the extension of this approach to the nonlinear setting. 
This is a purely mathematical problem, independent of the application field.
Using ideas from sheaf theory, we will extend in Section \ref{sec-equiv}
the projective-based approach above in order
to describe smooth planar curves as {\em classes of equivalent
functions}; here the equivalence notion will not rely on the one supporting
the construction of a projective space, as indicated for
(\ref{homogOhm}) above, but %somehow 
extends
this idea using the notion of {\em associate submersions}
%The difficulty met in this construction is related to the fact that
%we cannot assume a priori a common domain within $\R^2$ for
%all possible defining functions of characteristic curves, in clear %blatant
%contrast to the linear case since for the latter all linear forms
%are of course defined on the whole of $\R^2$. 
%Under the equivalence notion of ``being associates'' (a term borrowed 
(we borrow the ``associate'' term from the polynomial context).
This provides a nice setting to handle systematically
global implicit descriptions of planar curves.
Linearization at a given
point of the characteristic curve will naturally define a linear
form fitting the projective context mentioned above.
We believe these ideas %in Section \ref{sec-equiv} 
to be of independent
mathematical interest and %to be 
of potential use 
%not only in  nonlinear circuit theory
%but also 
in different branches of applied mathematics, science and engineering.

The formalism above can be naturally combined with 
determinantal, graph-theoretic tools  \cite{chaiken78, chen} to address in broad generality
certain qualitative aspects of nonlinear circuits. This is the second goal of
the paper, which is tackled in Sections \ref{sec-cir} and \ref{sec-bif}.
Following the philosophy of \cite{saddlenode, tbwp}, but using
very different tools, we will characterize in graph-theoretic terms
%and great generality
different analytical properties of nonlinear circuits, involving e.g.\ the unique solvability
of nonlinear resistive circuits or certain bifurcations of equilibria
in a dynamic context. 
We will also derive a 
general expression for the characteristic polynomial
of a nonlinear circuit at equilibrium,
in terms of the structure of spanning trees. All these results
hold under a fully-implicit form for resistors (and memristors, 
when present).
A key result in our approach, proved in \cite{homoglin}, is 
a projectively-weighted
version of the matrix-tree theorem. This  result
 makes it possible to exploit the homogeneous form
of the incremental resistances and memristances to address
 in an implicit context the problems mentioned above, without
the need to resort to (local or global) explicit descriptions of
the corresponding devices.
%will be used at different points in the analysis.
We consider in more detail a bifurcation problem in
Section \ref{sec-bif} where, specifically, we use our framework
to provide a graph-theoretic characterization of so-called
simple stationary bifurcation points in nonlinear circuits.
Finally, the reader can find some concluding remarks in Section \ref{sec-con}.
Throughout the paper, several examples illustrate the results.

%\section{Motivation: the linear setting}
%\label{sec-bac}

%Relevant already in the linear setting: e.g. assume $i=Gv$, $G$ the conductance matrix... 
%Then $Ai=0$ (needless to say, we choose the simplest possible context, disregarding sources which would 
%be accommodated in the right-hand side, etc., to focus
%on the key mathematical aspect of the problem)
%yields $AGv=0$ and, in terms of node potentials, $AGA^Te=0$. The nonsingularity of the coefficient
%matrix (characterizing the unique solvability of the circuit equations) involves a weighted
%Laplacian matrix and the classical form of the weighted matrix-tree theorem [REF] applies.

%But obviously a loss of generality; this cannot be solved by resorting to the dual description
%in current-controlled terms, nor even using hybrid descriptions in which some devices are 
%voltage-controlled and others are current-controlled. If we want to perform a full 
%study, comprising all possible parameter values
%(which is relevant from the perspective of bifurcation theory),
%any of the hypotheses above restrict the scope of the analysis.

%The key thing already at the device level: Ohm's law... change by...

\section{Associate submersions}
\label{sec-equiv}

In this section we address the following problem. From elementary linear algebra
we know that a straight line through the origin in the real plane can be described as 
the kernel of a nontrivial linear form, that is, a nonzero element of $L(\R^2, \R)$. Moreover,
all such linear forms can be identified as equivalent 
in terms of a standard relation, namely,
the one defining a projective space from a given vector space. Consider now the nonlinear
version of this problem: given 
a smooth curve in the plane, can we describe it as a zero set of a globally defined
function? Provided that this is the case (this actually
follows from well-known results in differential geometry), 
may all such functions be made
equivalent in some sense? We tackle this in what follows. By way of motivation,
we present in advance further details of the problem in the linear context in electrical circuit terms.

\subsection{Motivation: linear devices as points in a projective line}
\label{subsec-lindev}

%It is well-known in %elementary 
%projective geometry that straight lines through the origin
%in the real plane can be naturally identified with points in a %(so-called dual) 
%projective line. % projective space.
%The idea is that a straight line $px_1-qx_2=0$ (with $(p,q) \neq (0,0)$) can be seen as the common kernel of 
%a one-parameter family of linear forms defined on $\R^2$. The other way around, 
%two non-trivial forms $f$, $g \in L(\R^2, \R)$ are said to be equivalent
%in the standard construction of the projective space if
%there is a non-vanishing constant $\mu$ such that $f=\mu g$; from this relation it
%follows that both forms have the same kernel. The parameters $p, \ q$ above
%naturally define homogeneous coordinates of the projective point defined by the corresponding
%equivalence class.

%, a task which
%is undertaken in Section \ref{sec-equiv}, and then use it to address certain analytical
%problems in nonlinear circuit theory (Section \ref{sec-cir}). 
%Concluding remarks can be found in Section \ref{sec-con}.

%[BRIEFER, CHECK INTRODUCTION] Framework introduced in \cite{homoglin}: summarize. We focus on the case of linear resistors but the ideas apply
%to capacitors and inductors in entirely analogous terms. Homogeneous form of Oh%m's law (in the real setting;
%the results also hold in the complex domain):
%\begin{equation}\label{ohmh}
%pv-qi=0,
%\end{equation}
%where $(p,q)$ is a pair or real parameters not vanishing simultaneously.
%\neq (0,0)$.
%MAYBE BEGIN HERE:
As indicated in Section \ref{sec-intr},
following \cite{homoglin}
we look at %the solutions $(i,v) \in \R^2$ of \eqref{ohmh} 
the characteristic of a given linear electrical device
as the kernel of a linear form or, more precisely, as
the common kernel of all
linear forms from a certain equivalence class. Specifically,
in the vector space of linear forms 
$\R^2 \to \R$ (with the standard operations)
%better than ``from $\R^2$ to $\R$'' cause ``forms'' already includes codomain R
%(we will denote this dual space as $\left(\R^2\right)^{\hspace{-0.3mm}*}$) %=L(\R^2, \R)$ 
we identify two non-null forms $f$, $g$ as equivalent
if there exists a (non-zero) constant $\mu$ such that
\begin{equation}
  f=\mu g. \label{equiv}
\end{equation}
In the circuit context, a linear resistor will be a point
in the resulting 
quotient space, that is, {\em an equivalence class of linear forms}. 

This construction
defines a real projective line. %which we denote by $\PP(\left(\R^2\right)^{\hspace{-0.3mm}*})$.
Fix now the attention on a resistor defined by a characteristic such as
(\ref{homogOhm}) and let
$f(i, v)$ stand for the left-hand side of this equation.
By means of the basis defined by $f_1(i, v)=v$ and $f_2(i, v)=-i$
we may understand the parameters $p$, $q$ to define
homogeneous coordinates in the aforementioned projective line, 
since $(p:q)$ is the %homogeneous
pair 
for which the identity $f(i, v)=pf_1(i, v)+qf_2(i, v)$ holds. 
%(cf.\ . 
If we multiply the left-hand side of (\ref{homogOhm}) by a non-null constant, 
this constant arises as a common factor to both parameters $p$ and $q$, so that the ratio
is the same: this makes the homogeneous nature of the pair clear.
Such a coordinate pair $(p:q)$ is called
a {\em homogeneous description} of the resistance.

The advantage of this formalism is of course its generality.
  Regardless of the choice of a specific pair of homogeneous coordinates or, in other
terms, of the choice of a particular linear form within the equivalence class defining the resistor,
the (say, classical) resistance $q/p$ and the conductance $p/q$ are well-defined and unique
%(and each one of them is unique)
in the so-called
current-controlled and voltage-controlled affine patches 
defined by the conditions $p \neq 0$ and $q \neq 0$, respectively. 
The point
is that by using the projective
formalism we do not need to restrict the analysis to such patches, and 
this idea may
in principle apply throughout the whole of linear circuit theory.
The same holds for the impedance and the admittance in the complex domain.
A detailed discussion, focusing on the formulation of reduced models, can be found in 
 \cite{homoglin}. We refer the reader to 
\cite{OtroBryantProjective, chaikenArxiv, penin, smith72}
for some related approaches.
% (for general background on projective geometry see e.g.\ \cite{jeangallier,karensmith}).

\subsection{Nonlinear devices as equivalence classes of associate submersions}
\label{subsec-nonlindev}

\subsubsection{Global description of planar curves}

%As indicated in the Introduction, o
Our first goal in this paper is 
to extend this modeling approach to the nonlinear context.
%*Using ideas from sheaf theory, we extend in this Section
%the projective-based approach of \cite{homoglin} in order
To do so, we introduce in what follows a formalism to describe smooth planar curves as classes of equivalent
functions; the equivalence notion will not rely on the one supporting
the projective construction above,
%as indicated for (\ref{homogOhm}) above, 
but %somehow 
extends
this idea using the notion of {\em associate submersions}
(recall that a submersion is a smooth map with a surjective differential everywhere).
%(we borrow the ``associate'' term from the polynomial context).
%The difficulty met in this construction is related to the fact that
%we cannot assume a priori a common domain within $\R^2$ for
%all possible defining functions of characteristic curves, in clear %blatant
%contrast to the linear case since for the latter all linear forms
%are of course defined on the whole of $\R^2$. 
%Under the equivalence notion of ``being associates'' (a term borrowed 
Linearization at a given
point of the curve will naturally define a linear
form fitting the projective context of subsection \ref{subsec-lindev}. 
We will exploit these notions in the electrical
circuit context in later sections
but, for the sake of generality,
in the remainder of this one we work with abstract 
curves and functions.

All maps and manifolds will be (often implicitly) assumed to be
$C^{\infty}$, so that ``smooth'' means
$C^{\infty}$, even if this could be relaxed at many points.
%when talking about functions or maps, ``smooth'' will mean $C^{\infty}$
%**Notation ${\cal M} \to $ ${\cal C}$
%To be precise,
%Our essential working assumption is that 
%the characteristic curve 
The attention is focused on 
%of each device will be assumed to define a 
smooth, connected %differentiable 
1-manifolds in $\R^2$, which for brevity will be termed %called
{\em smooth planar curves}. %throughout. %the document. 
We adopt the usual
notion for manifolds in Euclidean space, defining them as regular submanifolds of $\R^n$
by assuming the existence of adapted charts
(find this e.g.\ in \cite[Definition 9.1]{tu}).
% or right at the beginning
%of Chapter 5 of \cite{spivak} **or in the Appendix**???.
%Essentially, t
This %ir our setting this point of view 
amounts to requiring that each point of 
the curve, with the topology inherited from $\R^2$, has a 
neighborhood diffeomorphic to a real interval \cite{lee}. %, milnor}.
%;
%this excludes, for example, the closed topologist' sine curve \cite[\S 9, example 9.3]{tu}.
%This way we avoid the topologist sine curve phenomenon, see Tu ex 9.3 or Boothby ex 4.10: the important point is that this guarantees
%the local level set theorem
%The sets ${\cal C}$ 
%For terminological brevity, we will say that such a characteristic defines a {\em smooth 
%planar curve}, this term denoting  a smooth, connected
%not necessarily closed  
%1-manifold in $\R^2$ throughout. 
We do not assume that the curves define
closed subsets of $\R^2$.
%Needless to say, self-intersections of the curve are precluded 
%in this working scenario. 
%For simplicity, we assume all characteristics to be connected.

%*Somewhere: A {\em smooth device} will be a device whose characteristic is a smooth planar curve.

%** Appendix: The notion above excludes e.g.\ the closed topologist' sine curve (\cite{tu}[Example 9.3]); the 
%important point is that regular...

%\subsubsection{Global implicit description of smooth planar curves}

Regular submanifolds are well-known to
%always 
admit local descriptions as
level sets of submersions
%Locally, such a description is always feasible for a manifold
%of arbitrary dimension in $\R^n$:
%This is a straightforward consequence of the implicit function theorem 
(see %and an explicit statement can be found e.g.\ in 
\cite[Proposition 5.16]{lee}). %Springer 2013 pdf, ok
%\cite[Proposition 8.12]{lee}. Old Springer 2003- %Google Books, 
%For smooth planar curves, this says that for every $x \in {\cal C}$ there exists
%a neighborhood $U \subseteq \R^2$ such that ${\cal C} \cap U$ is the zero set 
%of a submersion $g: U \to \R$. 
%In other words, locally around any point of ${\cal C}$, the 
%curve admits an implicit definition of the form $g(x)=0$.
%can be described by an equation of the form %[*maybe $(v,i)$*]
%\begin{equation} \label{implicit0}
%g(x)=0.
%\end{equation}
%(\ref{implicit}), in the understanding that
%$g$ is defined only on some neighborhood of the point (arbitrary dimension). 
%Such a locally defined 
%submersion $g$ is called a {\em local defining function} \cite{lee}. 
%We emphasize that, by construction,  either in this local sense or in the global one below,
%a defining function does not display any zeros away from ${\cal M}$.
Our first remark is that for planar curves the same kind
of description is also feasible in a global sense.

%\vs

\begin{propo}
Any smooth planar curve admits
a global description as the zero set of a smooth submersion
defined on some open subset of \hspace{0.5mm}$\R^2$.\vv
\end{propo}

\noindent %{\bf Proof.}
% but maybe not on the whole of $\R^2$. 
Even if it is not easy to find an explicit statement of this claim in the literature,
it is a straightforward combination of two well-known facts
in differential geometry, 
namely: that an orientable codimension-one manifold (hypersurface) 
in $\R^n$ is the zero set of a %smooth 
submersion $U \to \R$ for some open set $U$ including
the hypersurface (an explicit statement can be found for instance in \cite{spivak}),
%problem 5.14, p. 121, partitions of unity
and that 1-dimensional manifolds are orientable (see e.g.\ 
\cite{lee}). %Problem 10.1 in the 2003 version, pdf OLD
%Problem 15.13 in 2013 version

%From now on we may then assume that, in the general case, %the description
%(\ref{implicit0}) describes the characteristic 
This means that an implicit description $f(x)=0$ of a smooth planar curve ${\cal C}$
holds in a global sense, 
with $f$ defined on some open neighborhood $U$
of ${\cal C}$; %the curve; 
such a function $f: U\to \R$ 
is called a {\em global defining function} for ${\cal C}$. 
%\cite{lee}.
%Recall that we do not require the curve to define a closed set in $\R^2$, 
%%%%(Whitney lemma in Moerdijk Reyes 86; smooth Tiezte in MR86). However, 
%but the reader should be aware of the fact that, in this broader context, 
%the submersion might not be defined on the whole of 
Note that in general $U$ need not be the whole plane $\R^2$,
%: for a logarithmic spiral, 
%%different from the Archimedean spiral which reaches the origin, I think... check back
%which as a smooth planar curve in the sense above, the
%origin is in the closure of the curve and therefore any submersion defined on $\R^2$ 
%and vanishing on the curve would also vanish at the origin
%%origin as a limit point not belonging to the curve, which as a smooth planar curve
%%in the sense above 
although
%(by contrast, a %{\em compact} %connected
if the %regular planar 
curve
%1-manifold %without boundary 
%--make sure that this rules out the cases $[a,b)$ and $[a,b]$ in the classification theorem--/
%planar 1-manifold 
%${\cal M}$ 
is diffeomorphic to $\mathbb{S}^1$ %(or S alone, check notation everywhere) 
%--maybe better after classification theorem--
then it can indeed be described as the zero set of a submersion defined on 
%the whole of 
$\R^2$, as a consequence of the smooth Jordan-Sch\"oenflies theorem.
%\cite{carstenThomassen}. %% careful, this only topological, 
%which
%guarantees that there is a diffeomorphism $\R^2 \to \R^2$ mapping ${\cal C}$ onto $\mathbb{S}^1$).
%Therefore, f

\subsubsection{Equivalence classes of associate submersions} 
%A key remark arises here: 
%%[[(maybe for the Intro, first draft of the paragraph below)
%%The point now is, 
%if we think of a smooth nonlinear resistor as a (regular, in the sense above)
%set of points within the $(i, v)$-plane, this set admits 
It is clear that 
%any smooth planar curve admits infinitely many different global defining functions; that is, 
there are infinitely many submersions defining a given smooth  
curve ${\cal C}$.
%but want to describe
%it as a zero set, which g should be choose? More precisely, we 
Our present purpose %first goal %at this point 
is
%We would like 
to provide a formal definition making all of them equivalent
and extending the aforementioned projective-based
construction, which only applies to
%of the projective space 
%holding in 
linear problems.
%which holds in the linear context.
%making all such submersions equivalent in
%some sense and without the need to fix the set ${\cal C}$ a priori.
% much as in the linear case a point in a projective
%line comprises as equivalent all possible linear forms 
%with a prescribed kernel. Moreover, we want to do it in a way
%that particularizes/fits well this construction in the linear
%case when looking at the differential at a given point
%of the curve (which defines the linearized resistor)...]]
%%%%%%%
%The point now is, can we identify in some sense all the submersions that define the same zero set? Note that in the linear...
%In the linear case we achieved this by means of an equivalence 
%notion in the %vector 
%space 
%of %
%non-null %planar 
%linear forms, namely the equivalence notion (\ref{equiv}) 
%which constructs
%a projective space from a given vector space.
% %the equivalence notion supporting the projective space makes it possible to identify
%%essentially because this notion identifies as equivalent
%%all (non-zero) forms with the same kernel. 
%%In the nonlinear setting we want to do it in a way
%%that particularizes/fits well this construction in the linear
%%case when looking at the differential at a given point
%%of the curve (which defines the linearized resistor) [redactar].
The difficulty in the nonlinear case relies on the fact that 
the submersions eventually defining a zero set do not need to share a common domain, contrary to linear
forms which are well-defined on the whole of $\R^2$;
%$\R^n$ ($\R^2$ in the planar case); 
moreover, since
%if
%the curve 
${\cal C}$ is not fixed a priori there is no chance to localize the definition
on a neighborhood of the curve, as it is done for instance to define
germs. % or multigerms.
%The notion of a 
{\em Sheaves} %does the job, %performs the task, %
naturally arise here, 
since this notion accommodates families of functions
defined on open subsets of a given topological space without restricting a priori a 
common domain for all of them.
%of functions.
%The result above shows in essence that any planar 1-manifold ${\cal M}$
%can be mapped in a one-to-one manner to the set of submersions that define
%it as a zero set, each one defined on some open neighborhood
%of ${\cal M}$ and in a way such that any two of them are related by a
%nowhere zero factor in a (possibly smaller) open neighborhood
%of ${\cal M}$. To make this precise 
%and contained in the intersection of the domains of both functions.
%we make use of the {\em sheaf of submersions} defined on open sets of 
%$\R^2$. 
We refer the reader to \cite{hartshorne}
%, wedhorn} %CHECK BREDON and TENNISON
for general 
background on sheaf theory. For our purposes it is enough to 
consider the {\em sheaf of submersions} ${\cal S}$
%${\cal S}(\R^2, \R)$ %%%MAYBE SIMPLY ${\cal S}$ %or ${\cal S}(\R^2)$
(which is a sheaf of sets, cf.\ \cite{hartshorne})
%David Ayala, Geometric cobordism categories
as the family of all pairs $(f, U)$ with $U$ an open set
of $\R^2$ %(with the Euclidean topology) 
and $f: U \to \R$ a smooth %$C^{\infty}$ 
%(this assumption will be made throughout without further explicit mention) 
submersion. 
%For any open set, we denote as $S(U)$ the set of submersions 
%with domain $U$, which is a subset of the ring $C^{\infty}(U)$ of smooth functions defined on $U$. 
%%(for general sheaves, this set is called the set of {\em sections} over $U$).
%** Check if we use both... USE ${\cal S}$ for the whole sheaf and $S$ for submersions on a given set (section%s).**
%
%
%This notion .
%Needless to say, many different submersions may define the same zero set (e.g.\ enlarging the domain
%of a given submersion when possible, or even changing the function without affecting the zero
%set). 
In this context, the following definition %will make it possible 
%[capture the equivalence of all these functions/] identify all such cases 
%into a single one -allowing for 
will capture the sought correspondence between smooth planar 
curves and classes of global defining functions
(cf.\ Theorem \ref{th-associates}).

\vspace{1mm}

\begin{defin} \label{defin-associates}
Two smooth submersions %(or two arbitrary functions?) 
$f_1$, $f_2$,
%$(f_1, U_1)$, $(f_2, U_2)$ 
defined respectively 
on %two %arbitrary 
open subsets $U_1$, $U_2$ of $\R^2$, %/$\R^n$ a topological space,
are said to be {\em associates} if the following two conditions hold.

\vs

1. If $U_1 \neq U_2$, then $f_i \neq 0$ on $U_i - U_j$, for $\{i,j\}=\{1,2\}$.

\vs

2. If $U = U_1 \cap U_2 \neq \emptyset$, then 
there exists a nowhere zero smooth %$C^{\infty}$ 
function $\gamma: U \to \R$ 
%not vanishing at any point of $U$ and 
such that %(\ref{nz-equiv}) [there we denote f, g] 
$f_1 = \gamma f_2$
holds on 
the whole of $U$.
\end{defin}

\vspace{1.5mm}

%\noindent %We show in the Appendix that this defines an equivalence relation in the sheaf ${\cal S}$.
%Here we are borrowing the ``associates'' term from the polynomial setting (cf.\ \cite{hall}).
%%; we refer the reader to the Appendix for further remarks in this direction). 
%Note that in Definition \ref{defin-associates} and in
%Theorem \ref{th-associates} below
%we restrict the statement to the $(\R^2, \R)$-setting because our interest is just focused
%on planar curves.

%\begin{propo}
%As shown in Propo*** Appendix..., this is an equivalence relation... in the sheaf 
%((of planar submersions)) 
%${\cal S}$. %${\cal S}(\R^2)$.

%\end{propo}

%(This propo, or at least the proof, to the Appendix)

%\vs

\begin{propo} \label{propo-equivrelation}
The relation of being associates is an equivalence relation in the
sheaf ${\cal S}$.
\end{propo}

\vspace{1mm}

\noindent {\bf Proof.} 
The fact that the relation is reflexive and symmetric 
is clear. We need to show that it is also transitive. 
To this end, denote the relation by $\sim$ and 
suppose $(f_1, U_1) \sim (f_2, U_2)$ and 
$(f_2, U_2) \sim (f_3, U_3)$. We prove in what
follows that $(f_1, U_1) \sim (f_3, U_3)$.

We first need to check that $f_1 \neq 0$ on $U_1 - U_3$ (for $f_3$ the reasoning
is symmetrical). Split this set as
$U_1 - U_3 = [U_1-(U_2 \cup U_3) ] \cup [(U_1 \cap U_2) - U_3].$ 
The first set in the right-hand side is a subset of $U_1-U_2$ and
therefore $f_1 \neq 0$ there. The second one is a subset of 
$U_1 \cap U_2$, where the identity $f_1 = \gamma_{12} f_2$ holds
for some nowhere zero
function $\gamma_{12}$, and it is also a subset of 
$U_2 - U_3$, where $f_2$ is known not to vanish: we conclude that
$f_1=\gamma_{12} f_2$ is not zero there, either, and the first part
of the proof is complete.

We also need to show that if $U_1 \cap U_3$ is non-empty, then
there exists a nowhere zero, smooth function $\gamma_{13}$ 
such that $f_1 =\gamma_{13}f_3$
there. Write  
$U_1 \cap U_3 = [U_1 \cap U_2 \cap U_3 ] \cup [(U_1 \cap U_3) - U_2]$ 
and note that on the first we set may write $f_1 =\gamma_{12}\gamma_{23}f_3$,
where $\gamma_{23}$ is nowhere zero and
such that $f_2 = \gamma_{23}f_3$ on $U_2 \cap U_3$.
Let us then define
\begin{equation}
\gamma_{13}(x)= \begin{cases} 
\gamma_{12}(x)\gamma_{23}(x) & \text{ if } x \in U_1 \cap U_2 \cap U_3  \vs\\
f_1(x)/f_3(x)  & \text{ if } x \in (U_1 \cap U_3) - U_2.
\end{cases}
\end{equation}
Be aware that the quotient here is well-defined because $f_3$ does
not vanish on $U_3 - U_2$, which includes $(U_1 \cap U_3) - U_2$.
Note also that the identity  $f_1 =\gamma_{13}f_3$ holds trivially
in light of the definition of $\gamma_{13}$.
The function $\gamma_{13}$ so defined does not vanish because, on the one hand,
neither $\gamma_{12}$ nor $\gamma_{23}$ do 
on $U_1 \cap U_2 \cap U_3$
and, on the other,
$f_1$ is non-null on $(U_1 \cap U_3) - U_2 \subseteq U_1 - U_2$.
Finally, $\gamma_{13}$ is smooth; this is clear for any
$x \in U_1 \cap U_2 \cap U_3$ because this is an open
set and $\gamma_{13}$ is a product of smooth functions
there; for $x \in (U_1 \cap U_3) - U_2$, smoothness
follows from the fact that $f_3$ does not vanish on this set, and
by construction $\gamma_{13}$ equals 
the smooth quotient $f_1/f_3$ on some open neighborhood of $x$.
%both of which
%are defined on the open set $U_1 \cap U_3 \supseteq (U_1 \cap U_3) - U_2$. 
%
%the product $\gamma_{12}\gamma_{23}$ does not vanish on that 
%intersection and equals $f_1/f_3$ on the (open dense) subset where
%$f_3$ is not zero {\bf --here we are using that $f_3$ is a submersion; 
%there should be a lemma with this and what follows (cf also \eqref{eqcases}). 
%Think,
%maybe we don't need dense, ie maybe don't need a submersion. 
%Work with a neighborhood of the zero set-- PDTE}. 
%On the second
%set $f_3$ dos not vanish (since it is contained on $U_3 - U_2$) and
%therefore the quotient $f_1/f_3$ is well-defined and smooth there. 
%Glue both (with the aforementioned lemma) and we are done.
\hfill $\Box$

\begin{theor} \label{th-associates}
Two smooth submersions from the %above defined 
sheaf ${\cal S}$
defined above yield the same zero set if and only
if they are associates. %[Write better?... restrict planar... ${\cal S}(\R^2)$]
\end{theor}

%This was Proposition \ref{propo-associates} in the appendix. 

%\vv

\noindent {\bf Proof.} The ``if'' part of the claim directly follows from 
Definition \ref{defin-associates}. % above. We therefore focus on the 
For the ``only if'' part, what we need to show is that, if
$f_1,$ $f_2$ %(*or $f$, $g$) 
are two submersions with open domains
$U_1$, $U_2$ and defining the same zero set ${\cal C}$, there
exists %an open neighborhood $U \subseteq U_1 \cap U_2$ of ${\cal M}$ and 
a nowhere zero smooth function $\gamma: U \to \R$ 
defined on the intersection $U=U_1 \cap U_2$ 
such that
\begin{equation} \label{nzequiv}
f_1 = \gamma f_2
%$$f(v,i)=\kappa(v,i)g(v,i)$$
\end{equation}
%for all $(v,i) \in \tilde{U}$.
on $U$.
%\end{propo}
%\noindent {\bf Proof.} 
To do this, let us first fix $x^* \in {\cal C}$. Since by 
definition $f_2$ is a submersion, the differential $(df_2)_{x^*}$ is surjective
and this implies that at least one of the partial derivatives
${f_2}_{x_i}$ does not vanish (for brevity, 
we write $x_i$ as a subscript to denote partial differentiation). Assume w.l.o.g.\ that
the first partial derivative ${f_2}_{x_1} $ is not zero at $x^*$, and
define a local coordinate change by
the local diffeomorphism $y=\alpha(x)$ given by
$y_1=f_2(x)$, $y_2=x_2$.
%\begin{eqnarray*}
%y_1 & = & g(x) \\
%y_2 & = & x_2.
%\end{eqnarray*}
Denote the inverse of $\alpha$ as $\beta$, so that $x=\beta(y)$. Then
we have $f_2(\beta(y))=y_1$ and, by the hypothesis that 
$f_1$ and $f_2$ define the same zero set,
we have (always in a local sense) $f_1(\beta(y))=0$ iff
$y_1=0$. According to Hadamard's lemma (see e.g.\ \cite{giaquinta}), it follows
that $f_1(\beta(y))$ can be written as 
$f_1(\beta(y))=\kappa(y)y_1$
for some locally defined function $\kappa$. In $x$-coordinates the latter
relation reads as
$f_1(x)=\kappa(\alpha(x))f_2(x),$
so that $\kappa \circ \alpha$ yields a
locally defined function, say $\gamma^{x^*}$, for which the identity 
$f_1(x)=\gamma^{x^*}(x)f_2(x)$ holds on some neighborhood of $x^*$.

To check that $\gamma^{x^*}$ does not vanish at $x^*$ (and hence
on a sufficiently small neighborhood of this point) it is enough
to use $f_2(x^*)=0$ 
in order to derive
%\begin{equation} %\label{differential1}
$(df_1)_{x^*} = \gamma^{x^*}(x^*) (df_2)_{x^*}$
%\end{equation}
%\begin{equation} \label{differential1}
%f_v(i^*, v^*) = \kappa(i^*, v^*) g_v(i^*, v^*), \ \ \
%f_i(i^*, v^*) = \kappa(i^*, v^*) g_i(i^*, v^*)
%\end{equation}
by Leibniz's rule. The fact that $df_1$ is surjective at $x^*$
%\neq g'(i^*, v^*)$
implies $\gamma^{x^*}(x^*) \neq 0$, as claimed.
%In the understanding that the domain of
%$\gamma$ is restricted if necessary 
%as to satisfy all the requirements above, denote
%this domain by $U^{x^*}$.

%It remains to show that gluing together all the functions $\gamma$
%defined on the open sets $U^{x}$ above we get a well-defined global
%function. We define $U$ as the union $\cup_{x \in {\cal M}}U^{x}$
%and make use of the fact that the set of nowhere zero functions on $U$ define
%a sheaf on $U$ \cite{arapura}. Since by
%construction $\gamma$ is uniquely defined on non-empty intersections
%$U^{x_1} \cap U^{x_2}$ for $x_1,$ $x_2 \in {\cal M}$
%it follows that, gluing together all 
%the aforementioned local functions, we get a 
%well-defined function $\gamma: U \to \R$. 

%SIMPLER: local $\gamma$
%equals the quotient $f(x)/g(x)$ away from ${\cal M}$ and
%extends smoothly to ${\cal M}$. Write as: $\gamma^{x^*}(x)$ the map
%above on the set $U^{x^*}$. Then the map
To finish the proof, we define $\gamma$ globally on the intersection $U=U_1 \cap U_2$ as
\begin{equation} \label{eqcases}
\gamma(x) = \begin{cases} 
f_1(x)/f_2(x) & \text{ if } f_2(x)\neq 0 \\
\gamma^{x}(x) & \text{ if } f_2(x) = 0,
\end{cases}
\end{equation}
%is well-defined, everywhere non-null and smooth (or use
%a covering of ${\cal M}$. Well-defined
%because it is so on the open dense subset where
%$g \neq 0$... / ...
%since its value at any $x$ (or in the domain intersections...)
%is uniquely defined 
%${\cal M}$ is on the closure
%of $g \neq 0$ and hence equals the limit... (by continuity). 
which obviously does not vanish and meets the requirement depicted in (\ref{nzequiv}) %$f=\gamma g$.
(recall that we are assuming $f_1$ and $f_2$ to define the same zero set).
One can easily check that $\gamma$ is 
smooth because, for any $x^* \in {\cal C}$, the function $\gamma(x)$ equals 
the locally defined one $\gamma^{x^*}(x)$
on some neighborhood of $x^*$; indeed, for any $x \in {\cal C}$ sufficiently close to $x^*$, the prescribed value
 $\gamma^x(x)$ must equal $\gamma^{x^*}(x)$ because $\gamma^x$ and $\gamma^{x^*}$
coincide, by construction, 
in the open and dense subset %of the domain of $\gamma^{x^*}$ NO, $\gamma^x$ also locally defined
defined by the condition 
$f_2(x) \neq 0$ on the intersection of their domains. % intersection.
%in the open and dense subset  of their domains intersection
%%of the intersection of their domains
%defined by %the condition 
%$g(x) \neq 0$.
\hfill $\Box$

\vs

\noindent Theorem \ref{th-associates} yields a one-to-one
correspondence between smooth planar curves and equivalence classes of global defining functions,
%(note that the relation of being associates is indeed an equivalence relation, as shown in Proposition \ref{propo-equivrelation} in the Appendix). 
since we have shown that two submersions $f_1$, $f_2$ 
define the same zero set if {\em and only if} a relation of the form $f_1 =  \gamma f_2$ holds
for some nowhere-zero function $\gamma$ defined
on an open neighborhood of this zero set. 
%Mind the similarity with the linear case, in which the equivalence 
%between two linear forms defining the same kernel amounts to the existence of a non-null multiplicative constant relating both.
%The reader is referred to the Appendix for further remarks in this direction. 
%\noindent 

\subsection{Linearization}
\label{subsec-lin}

Linearization at a given point naturally amounts to the projective
construction of subsection \ref{subsec-lindev}.
Indeed, fix any submersion $f_1$ defining a smooth planar curve ${\cal C}$. % globally. 
The differential of $f_1$ %any of these submersions, say $g$, 
at any given $x \in {\cal C}$
defines the linear form $(df_1)_x: \R^2 \to \R$ (mind the canonical
identification $T_x \R^2 \sim \R^2$). Following the construction 
in Theorem \ref{th-associates}
(cf.\ \eqref{nzequiv}), and using
%use of 
the property $f_2(x)=0$ for $x \in {\cal C}$, $f_2$ being any other 
submersion defining ${\cal C}$, we derive
%\begin{equation} \label{differs}
$(df_1)_x=\gamma(x) (df_2)_x$
%\end{equation}
 with $\gamma(x) \neq 0$:
therefore, 
$(df_1)_x$ and $(df_2)_x$ are equivalent linear
forms in the projective sense indicated in subsection \ref{subsec-lindev}.
%Section \ref{sec-intr}.
%in (\ref{equiv}) 
%and
%so they define the same point in the projective space $\PP(\left(\R^2\right)^{\hspace{-0.3mm}*})$. 
Geometrically, the remarks above simply express that the common kernel of all these differentials 
%describes
characterizes
the tangent space to %${\cal M}$ 
the curve at $x$. %, within $T_x \R^2 \sim \R^2$. 

It is worth remarking, however, that the connection with the 
projective construction in the linear context is much deeper, and
we 
%We 
finish this section with a remark in this direction.
%on the sense in which this construction
%generalizes the
%aforementioned projective construction. 
Fix a smooth planar curve
${\cal C}$ and %restrict the attention to open subsets of $\R^2$ which include ${\cal C}$.
%Consider, specifically, 
consider two open 
neighborhoods $U_1$, $U_2$
of ${\cal C}$ and two smooth functions $f_i: U_i \to \R$, $i=1,2$.
The pairs $(f_1, U_1)$ and $(f_2, U_2)$ are {\em germ-equivalent}
%(not in the ``associates'' sense above) 
at ${\cal C}$
if there exists an open set $U \subseteq U_1 \cap U_2$ containing ${\cal C}$
where $f_1$ and $f_2$ coincide. A {\em germ at ${\cal C}$} is an equivalence
class of such pairs. 
%**(mention multigerm stuff if not before), Osaki*. 
%Somewhere, maybe: note that the associates equivalence in section *** is defined without the need to introduce
%germs, whereas here we make use of them to construct an algebraic structure;
%both constructions are compatible, however, because
%the associates relation is coarser than the germ-equivalence CAREFUL AWAY FROM THE MANIFOLD, CHECK.
The set of germs at ${\cal C}$, denoted
by $G({\cal C})$, %--
%Note additionally that the addition and the product of $f_1$ and $f_2$
%are well-defined on the open neighborhood $U_1 \cap U_2$ of ${\cal C}$. Moreover,
%these operations are 
%easily proved compatible with the germ-equivalence relation. This way
%we are therefore naturally
%led to a (commutative and unital) ring structure on $G({\cal C})$.
%In this ring, one can check that the set of germs of smooth functions vanishing on 
%${\cal C}$ (to be denoted by 
%$C^{\infty}({\cal C},0)$) defines an ideal. 
%, that is, an additive subgroup  which is closed under multiplication by any element of the ring. 
%%%%%%%%%%% This provides additional info but not really necessary
%In particular, the germs of submersions
%which vanish on ${\cal C}$ (to be denoted by $S^{\infty}({\cal C},0)$)
%belong to this ideal; moreover, the reasoning in
%Proposition \ref{propo-associates} above essentially shows that this ideal 
%(hence a so-called principal ideal) 
%is actually generated by any submersion (vanishing on ${\cal C}$, so better later): this is a known result in the differential geometry literature, %, although it is often used only in a local sense: 
%see e.g.\ \cite[Lemma 2.1]{moerdijk}. MAYBE THIS REMARK EARLIER OR EVEN IN THE PROOF
%.. More... Parshin pp 49-50 QUOTIENT SHEAF. Somewhere "divide each other"
%Maybe mention VARIETY POINT OF VIEW. Note that in any case our simple context makes it worth a self-contained proof. DON'T REALLY NEED ALL THIS.
%%%%%%%%%%
%The set $G({\cal C})$ 
has a natural (commutative and unital) ring structure.
%Now, denote by 
Now, the set $N({\cal C})$ 
of germs of smooth
functions not vanishing at any point %/on a neighborhood 
of ${\cal C}$ %, to be denoted by . 
%This 
is %actually 
the %note that $N^{\infty}({\cal C})$ is the 
multiplicative group of units (elements with a multiplicative inverse)
in %the ring 
$G({\cal C})$.
%, namely, %that is, the set of 
%the . 
This set makes it possible to stress the similarities with the construction
in the linear setting reported %above: %
in subsection \ref{subsec-lindev}: 
indeed,
two linear forms are equivalent in the projective construction,
%defined by (\ref{equiv}) 
and yield the same kernel,
iff %and only if 
they differ by a non-vanishing constant, these
constants being the units in the ring of polynomials. %In our construction and  a
According to Definition \ref{defin-associates},
two submersions are equivalent, and define the same zero set ${\cal C}$,
%, cf.\ Theorem \ref{th-associates}) 
iff %and only if 
they differ by a function which is a representative of a germ in the set of units
$N({\cal C})$.
%on the intersection of their domains, 
%and, as indicated above, $N({\cal C})$ 
%defines the %nowhere zero functions 
%set of units in the ring $G({\cal C})$. 
%Note also that t
This analogy 
also
supports borrowing the ``associates'' term from the polynomial context,
since two polynomials are associates \cite{hall} if 
they differ by a non-null multiplicative constant (again, a unit in the ring of polynomials), 
whereas two submersions
are associates if they differ by a nowhere zero function (that is, a function yielding a unit in $G({\cal C}$)) on a neighborhood of 
their zero set ${\cal C}$.

%maybe notation $S_0$ for submersions with a non-vanishing zero set... that is $S-N$... and
%$L_0$ regular linear forms (which are submersions)

%$$S_0/\sim$$

%and

%$$L_0/\approx$$

%where the equivalences indicated by ${\sim}$ and $\approx$ correspond to the existence of a "unit" multiplicative factor (write) in both families, nowhere zero functions in the nonlinear case, constants in the linear case.

%Close with a pointer to section "nonlinear resistors... linearization"...

\section{%Qualitative properties of n
Nonlinear circuits with implicit characteristics}
\label{sec-cir}

We now drive the attention to nonlinear circuit theory, with the focus
on problems which involve implicitly-defined characteristics. In this section we address
some general qualitative properties of such circuits (possibly including memristors),
combining the framework of Section \ref{sec-equiv}
with some determinantal, graph-theoretic tools which 
are compiled in subsection \ref{subsubsec-pwmtt}.
Section \ref{sec-bif} will be specifically focused on bifurcation properties.

\subsection{Nonlinear devices in electrical circuits. Homogeneous
incremental resistance}
\label{subsec-homogres}

%******** REMEMBER THE FORMER VERSION ON PURELY RESISTIVE CIRCUITS IS IN DIRECTORY AML
%\section{Nondegenerate solutions of nonlinear circuits}
%\label{sec-cir}
%\paragraph{Linearization.} 
%Fix any submersion $f_1$ defining a smooth planar curve ${\cal C}$. % globally. 
%The differential of $f_1$ %any of these submersions, say $g$, 
%at any given $x \in {\cal C}$
% etc.
Let us elaborate on the ideas discussed at the beginning of subsection \ref{subsec-lin}.
In that context, 
the standard coordinates $(x_1, x_2)$ on $\R^2$ yield two globally
well-defined linear forms $dx_1$ and $dx_2$ such that 
$dx_i(v)=v_i$ ($i=1, 2$) for any vector $v=(v_1, v_2)$. %at any $x$
%: these two forms define the canonical basis of the dual space 
%$\left(\R^2\right)^{\hspace{-0.3mm}*}$.
%For later notational convenience, w
Now, at  
any point $x$ of a given smooth planar curve ${\cal C}$, and using again the 
the canonical
identification $T_x \R^2 \sim \R^2$, we choose $(dx_2, -dx_1)$
%******** change from dx1 -dx2 to  dx2, -dx1 for consistency with order i v in the circuit context****** 
%(simpler notation than $((dx^1)_x, -(dx^2)_x)$) 
as a basis for the cotangent space at $x$ (that is,
the space of linear forms defined on $T_x\R^2$).
In this basis, the differential at $x \in {\cal C}$ of any submersion $f$ 
defining ${\cal C}$ % \in S^{\infty}({\cal M},0)$ [OJO ESTA NOTACION]
has coordinates  
$\left(f_{x_2}, -f_{x_1}\right).$ 
Here, as above, we are using
 subindices to denote partial differentiation, and 
note that for notational simplicity we omit the dependence on $x$. 
Worth emphasizing is the fact
 that any other defining submersion $g$ may well yield different partial
 derivatives, but one can easily check
%it follows from (\ref{differs}) 
that $\left(g_{x_2}, -g_{x_1}\right)$ need be collinear
 with $\left(f_{x_2}, -f_{x_1}\right)$.
%$$\left(\frac{\partial f}{\partial x_1},  -\frac{\partial f}{\partial x_2}\right).$$
%[NOTACION PARA DERIVADAS PARCIALES...]
%The corresponding point in $\PP(\left(\R^2\right)^{\hspace{-0.3mm}*})$ 
%(that is, the linearized resistor) 
%is hence described by the
%homogeneous coordinates
%$(f_{x_1} : -f_{x_2})$ ++ORDER++, and i
It is then clear %from their homogeneous nature 
that 
 $(f_{x_2} : -f_{x_1})$ and $(g_{x_2} : -g_{x_1})$ define homogeneous coordinates of the same point,
regardless of the choice of the defining submersion.
%in the projective space $\PP(\left(\R^2\right)^{\hspace{-0.3mm}*})$. 
%, with $g$ standing for a second defining submersion, as above.
%(for notational simplicity we use here subindices to denote partial differentiation).
%$$\left(\frac{\partial f}{\partial x_1}: 
%-\frac{\partial f}{\partial x_2}\right).$$

%As indicated above, %by construction,
%the set of homogeneous
%coordinates are indeed independent of the actual choice of the submersion
%$f  \in  S^{\infty}({\cal M},0)$.
%[DE NUEVO NOTACION]
 %Already above

%\noindent {\bf Homogeneous resistance in the nonlinear context.}
The ideas above are geometric in nature and apply to any smooth curve %characteristic
in the plane, regardless of any actual physical meaning of the coordinates involved.
In what follows we focus on the case in which the coordinates of the Euclidean plane
are the current and the voltage in a given circuit branch, 
in order to emphasize the electrical meaning of the notions to 
be introduced. We therefore write
% and for better clarity 
%Let us now focus explicitly on the case in which the Euclidean plane is used to model 
%voltage-current pairs, so that
%let us write 
$x=(i, v)$ from now on.
%** $(v,i)$ or $(i,v)$**, 
%The framework above yields a one-to-one correspondence
%between (say, regular) %smooth 
%nonlinear resistors and equivalence classes of associate planar submersions.
Accordingly, ${\cal C}$ will now stand for the characteristic
of a nonlinear resistor in an electrical circuit. After fixing a point $x \in {\cal C}$,
the projective point defined by the 
equivalence class (in the sense of subsection \ref{subsec-lindev})
comprising the differentials of all defining submersions
is the {\em linearized resistor} at $x$. 

The following definition, which
naturally follows from the remarks above, makes it possible
to handle the incremental resistance notion
in a fully-implicit setting.

\begin{defin} \label{defin-homogresist}
Let a smooth planar curve\hspace{0.5mm} ${\cal C}$  be the characteristic of 
a nonlinear resistor. Assume that $f$ is any smooth submersion defining this characteristic, 
and fix $(i, v)  \in {\cal C}$. The
pair of homogeneous coordinates 
$$\left(\frac{\partial f}{\partial v}(i, v) : -\frac{\partial f}{\partial i}(i, v)\right)$$
is called  the {\em homogeneous incremental resistance} of the nonlinear resistor at $(i, v)$.
\end{defin}

%\vv

\noindent 
We emphasize that the homogeneous incremental resistance does not depend
on the choice of $f$, for the reasons indicated above,
%because homogeneous coordinates are defined only up to a scalar. More precisely and as explained above,
%the choice of any other submersion defining ${\cal C}$ yields at any point of the characteristic
%a collinear vector of partial
%derivatives and therefore defines the same projective point.
%In this context... (f or g)... any representative...
%the homogeneous coordinates $(f_v(i, v) : -f_i(i, v))$ define
%the {\em homogeneous incremental resistance} at $(i, v) \in {\cal M}$.
%For later use 
and that 
%our formalism 
%this notion 
%makes it possible to define 
this definition handles
the incremental resistance without resort to a %local
description %of the characteristic  
either in terms of the current $i$
or the voltage $v$.
%Now, since $f$ is a submersion, at least one of the partial derivatives above does not vanish.
%In particular, 
Locally, at least one of these descriptions is always feasible,
since at least one of the partial derivatives of the submersion $f$ does not vanish.
If the partial derivative $f_i(i, v)$ does not vanish 
at a given point then the implicit
function theorem  supports a local description 
%of the resistor 
%near this point 
in
the %current-
voltage-controlled form $i=g(v)$, %$v=\rho(i)$, 
with incremental conductance %(classical) resistance is
$g'(v)=-(f_i(g(v),v))^{-1}f_v(g(v),v).$
%$\rho'(i)=-(f_v(\rho(i),i))^{-1}f_i(\rho(i),i)$.
%where, again, we note that the ratio does not depend on the choice of $f$.
%In other words, 
This is actually a {\em dehomogenization} of the homogeneous
incremental resistance: indeed, in this setting an admissible choice for
the homogeneous coordinates above %defining the homogeneous incremental
%resistance 
is $(g': 1)=(-f_i^{-1}f_v:1)$, where we omit the dependence on $v$
for notational simplicity.
Dual remarks hold
%The same holds
for the (classical) 
incremental resistance.
%conductance.
% $\xi'(v)=-(f_i(v,\xi(v)))^{-1}f_v(v,\xi(v)),$
%which is well defined on regions where the partial derivative $f_i(i, v)$ does not vanish, allowing
%for a local voltage-controlled description $i=\xi(v)$ of the characteristic.
But note that the homogeneous formalism avoids the need
to perform any of these local reductions and makes it possible to handle
the incremental resistance in homogeneous form at all 
points of a fully-implicit 
characteristic. %, in the terms stated in Definition \ref{defin-homogresist}.
Remark finally that for a linear characteristic $pv-qi=0$ the notion
above yields the homogeneous resistance in the form $(p:q)$, 
the classical resistance and conductance being well-defined as $q/p$ and $p/q$
on the regions of the parameter space where the respective denominators
do not vanish, consistently with the framework of \cite{homoglin} (cf.\ subsection \ref{subsec-lindev} 
above).

%\paragraph{Passivity.} W
Finally, we define a device as {\em strictly locally passive} at a given point
$(i, v)$ of a smooth characteristic if for some (hence
any) defining submersion both $f_v(i, v)$ and $f_i(i, v)$ 
are non-vanishing and have opposite
signs. 
%Alternatively, if we write %Maybe write somewhere 
%$p(i, v)=f_v(i, v),$  $q(i, v)=-f_i(i, v)$ then both $p$ and $q$ must not vanish and
%have the same sign. One way or another, t
The idea behind this notion is that
the (classical) incremental resistance and the incremental conductance, as defined
above, are positive: note that the non-vanishing assumption for both derivatives 
in strictly locally passive regions guarantees that both current- and voltage-controlled descriptions are locally feasible. Many of the
determinantal-based
results to be discussed later become relevant in contexts in which
at least one device becomes non-passive on some operating region.

%%%%%KEEP NOTATION p for homogeneous (u-) description

%\

%*Somewhere mention sources

%\subsubsection
%\paragraph{Reactive devices.}
%%MAYBE ``OTHER DEVICES'', BRING SOURCES AND MENTION RESISTORS.}
%The same ideas apply in a straightforward manner
%to nonlinear capacitors, whose characteristics
%relate charge %(denoted by $\sigma$) 
%and voltage, and also for nonlinear
%inductors, defined by a characteristic relating magnetic flux
%%$\varphi$ 
%and current. We leave details to the reader.
%%; see in any case subsection ** parametric caps etc.

%\section{Concluding remarks}
%\label{sec-con}

%* We have extended in this paper the homogeneous approach of \cite{homoglin} to 
%uncoupled nonlinear electrical circuits, possibly including memristors. 
%* Using the formalism
%of sheaf theory, we have provided a construction 
%%of independent mathematical interest
%which identifies smooth planar curves (describing the characteristics of individual
%devices in the circuit context) with classes of equivalent submersions, much as in the linear setting
%the projective construction of \cite{homoglin} identifies linear devices with equivalent
%linear forms. 

%\section*{Appendix. On associate submersions}

%\mbox{} \vs

%\vspace{-3.6mm}

\subsection{%Nondegenerate solutions of nonlinear r
Nonlinear resistive circuits}
%Handling fully-implicit characteristics in %Unique solvability and DC operating points of 
%nonlinear circuit theory} 

We show in this and the following subsections the way
in which the formalism above can be used
 in nonlinear circuit modelling and analysis, specially
when one needs to employ 
fully-implicit descriptions of the devices. We begin with
resistive problems.
%Future research should illustrate further applications of this framework.
Even if this defines a non-dynamic context, these problems
will pave the way for a smooth introduction to some key tools extensively used later.
Since subindices will be henceforth used to distinguish digraph branches,
in what follows we resort to the conventional notation concerning partial derivatives.

\subsubsection{Explicit vs.\ implicit characteristics}
\label{discussion}

%- Dehomogenization, nodal, classical MTT
Let us consider a connected circuit with $n$ nodes and $m$ branches
composed of independent voltage/current sources and %(possibly nonlinear) 
smooth resistors without
%not displaying 
coupling effects. By letting
$i \in \R^m$, $v \in \R^m$ stand for the current/voltage vectors, 
the formalism above makes it possible to describe
the set of characteristics of the whole set of devices via a single map $$f: U \to \R^m$$
(with $U=U_1 \times \ldots \times U_m$, each $U_j$ open in $\R^2$), $f$ being a product of $m$ 
submersions $f_j: U_j \to \R$, that is,
$f(i, v)=\left(f_1(i_1, v_1), \ldots, f_m(i_m, v_m)\right)$. 
Every individual submersion $f_j$ is defined
up to an everywhere nonzero functional factor $\gamma_j$, as detailed in Section \ref{sec-equiv}.
Be aware of the fact that resistors and sources are treated in a unified manner;
e.g.\ a current source in the $j$-th branch is simply defined by an everywhere null derivative
$\partial f_j/ \partial v_j$.
%(this notation for partial derivatives  is intended
%to avoid a confusion in this context with subscripts corresponding to circuit branches). 

We assume that Kirchhoff laws are written as $Ai=0$, $Bv=0$ for appropriate matrices
$A \in \R^{(n-1)\times m}$, $B \in \R^{(m-n+1)\times m}$ (see e.g.\ \cite{bollobas, wsbook, homoglin}; additional details are given in 
subsection \ref{subsubsec-pwmtt} below). 
The circuit equations then read as
\begin{subequations}\label{reseq}
\begin{eqnarray} 
Ai & = & 0 \label{reseqa}\\
Bv & = & 0 \label{reseqb}\\
f(i, v) & = & 0.\label{reseqc}
\end{eqnarray}
\end{subequations}
This simple model already makes it possible to 
illustrate further a digression sketched
in Section \ref{sec-intr}, namely the
one concerning the feasibility of working with implicit 
descriptions throughout the whole circuit analysis.
Assume we want to analyze the 
isolation of a given solution by examining whether the matrix of partial
derivatives of the left-hand side of (\ref{reseq}), that is,
\begin{equation} \label{matrix0}
\begin{pmatrix} A & 0 \vspace{0.5mm}\\
0 & B \vspace{0.5mm}\\
\ \hspace{0mm}\displaystyle \frac{\partial  f}{\partial i} \ \hspace{3mm}& %(i,v) & 
\ \hspace{0mm} \displaystyle\frac{\partial  f}{\partial v} %(i,v)
\ \hspace{3mm}
\end{pmatrix}
\end{equation}
 has full rank at this solution.
A typical approach to this kind of problems involves a model reduction, 
%from order $2m$ (system (\ref{reseq}) involves $2m$ equations and %$2m$ 
%unknowns)
%to $m$, %Careful, this might naturally lead to homogeneous variables
%Add nodal reduction is to n-1
%. A typical way to reduce the model is to restrict the form of (\ref{reseqc}), for instance to 
which is based on imposing a global control structure on the devices.
For instance, one might assume that (\ref{reseqc})
can be globally recast in the voltage-controlled form
$i=g(v)$ 
%instead of the more general one depicted in ,
%describes the characteristics of all resistors
%exists for all $v$ lying on some $m$-dimensional domain 
(in particular, this voltage-control assumption 
entails that all sources are current sources, meaning
that 
%hold for 
the branches accommodating sources
are governed by relations of the form $i=i_s$; % with $i_s$ constant in our DC context, ; 
these are implicitly included in the map $g$ above).
Equation (\ref{reseqb}) may in turn be 
easily reduced in terms of an $(n-1)$-dimensional vector 
by recasting
Kirchhoff's voltage law as $v=A^{\tra}e$; we may assume for simplicity that $A$
is a reduced incidence matrix so that
%the reader may think of 
$e$ is the corresponding vector of node potentials. 
%, which is obtained when $A$ is chosen as a reduced incidence matrix. 
Altogether, this process reduces (\ref{reseq}) to the 
(so-called nodal) form
\begin{eqnarray} \label{reseqNodal}
Ag(A^{\tra}e)  =  0. 
\end{eqnarray}
To address the aforementioned problem 
in this restricted context, we may now differentiate the left-hand side of 
(\ref{reseqNodal}), so that the maximal rank condition %above 
relies on the nonsingularity
of  the weighted Laplacian
matrix 
\begin{eqnarray} \label{nodal}
A g'(A^{\tra}e)A^{\tra}.
\end{eqnarray}
The classical form of the weighted matrix-tree theorem  \cite{chaiken78} naturally applies
here. This theorem %Maurer 
says that the determinant
of (\ref{nodal}) equals %(maybe up to a nonzero constant, depending on the choice of $A$),
the sum of the products of tree incremental conductances $g_j'$ 
(this stands for the derivative of the $j$-th component of $g$; recall
that we exclude coupling effects among the different devices)
extended over the set of spanning trees. Note in particular that the derivative %defining
$g_j'$ vanishes at current sources and therefore the sum can be restricted to so-called
proper trees, which have all current sources in the cotree.

Our point is that the scope of this analysis is obviously 
restricted by the voltage-control assumption above.
Needless to say, a loss of generality also occurs under the dual current-control 
assumption, as well as in hybrid descriptions.
By contrast, we may keep the generality of the 
fully-implicit formalism for our present purpose by resorting to a form
of the matrix-tree theorem which does not impose (even locally) any restriction
on the controlling variables. This will naturally involve the 
homogeneous form of the incremental
resistance introduced in Definition \ref{defin-homogresist}, and will make it possible to assess directly the non-singularity
of the matrix (\ref{matrix0}) without assuming
any control structure on the devices. %resistors. %in general terms.
%of partial derivatives of (\ref{reseq}). 

\subsubsection{A projectively-weighted version of the matrix-tree theorem}
\label{subsubsec-pwmtt}

The result stated in Theorem \ref{th-pwmtt} is proved in \cite{homoglin}. It involves %an arbitrary choice
arbitrary choices of the reduced
cut and cycle matrices
$A \in \R^{(n-1) \times m}$ and $B \in \R^{(m-n+1)\times m}$ 
of a given (connected) digraph, making it possible 
to describe
Kirchhoff's current and voltage laws as $Ai=0$ and $Bv=0$, as in 
(\ref{reseq}) above. 
The reader is referred to \cite{bollobas} 
for background on such matrices and, more generally, on digraph theory.

We denote by ${\cal T}$ the set of spanning trees in the digraph, 
which for simplicity is assumed connected. Spanning trees
will
be described by the set of indices of their constituting branches (or
{\em twigs}),
so that $T$ and $\overline{T}$ stand for the indices of branches in a spanning tree $T$ and its
cotree, respectively. The branches indexed by $\overline{T}$ are
called {\em chords} or {\em links}.

\begin{theor} \label{th-pwmtt}
For any choice of the reduced cut and cycle matrices
$A$, $B$ of a given connected digraph, there exists a non-vanishing
constant $k_{AB}$ 
%Assume that $A$ and $B$ are two given reduced cut (or incidence) 
%and cycle matrices of a connected digraph.
%Let $P$, $Q$ be arbitrary diagonal matrices, with $p=(p_1, \ldots, p_m)$ and
%$q=(q_1, \ldots, q_m)$ the vectors of diagonal entries of $P$ and $Q$.
such that the identities
\begin{equation} \label{kir-hom}
\det 
\begin{pmatrix} A & 0  \\ 0 & B \\ -Q & P  \end{pmatrix}
=\det 
\begin{pmatrix} AP \\ BQ \end{pmatrix}
=k_{AB}\sum_{T \in {\cal T}} 
\left( \prod_{j \in T} p_j \prod_{k \in \overline{T}} q_k\right), 
\end{equation}
hold for any pair of diagonal matrices $P=\mathrm{diag}(p_1, \ldots, p_m)$, 
$Q=\mathrm{diag}(q_1, \ldots, q_m)$.
\end{theor}

\noindent We remark that it is always possible to choose $A$ and
$B$ so that $k_{AB}=1$;
%even if 
%for our purposes it is enough 
%we do not need to make use of this; 
as shown in \cite{homoglin},
the condition $k_{AB}=1$ is met 
in particular when $A$ and $B$ are the fundamental
cutset and cycle matrices defined by a spanning tree. Such fundamental
matrices take the form
\begin{equation} \label{fund}
A =
\begin{pmatrix} I & K    \end{pmatrix}, \ \ B =
\begin{pmatrix} -K^{\tra} & I \end{pmatrix}
\end{equation}
for a given matrix $K$.

The polynomial in the right-hand side of (\ref{kir-hom}) is 
the multihomogeneous form of the Kirchhoff 
polynomial or tree-enumerator polynomial. The ``multihomogeneous''
label %essentially
reflects the fact that every circuit branch $j$ contributes exactly one factor (either $p_j$
or $q_j$) to each monomial, so that the polynomial is homogeneous of degree one in each
pair of variables $(p_j, q_j)$, for $j=1, \ldots, m$. Specifically, the $j$-th branch contributes
the factor $p_j$ or $q_j$ to the monomial defined by a given spanning tree if the branch is a twig or a chord,
%belongs to the tree or to the cotree, 
respectively, for that tree.

%Dehomogenization (maybe not here but below): nodal**

\subsubsection{Nondegenerate solutions of resistive circuits}

Theorem \ref{th-pwmtt} makes it easy to characterize 
the nondegeneracy of
solutions of the circuit equations (\ref{reseq}) in fully-implicit form, in terms of the partial
derivatives which arise in the definition of the incremental resistance in
homogeneous form (cf.\ Definition \ref{defin-homogresist}).
We say that a given solution $(i^*, v^*)$ of (\ref{reseq}) is nondegenerate if 
the matrix (\ref{matrix0}), evaluated at that point, 
%of derivatives 
%\begin{equation} \label{matrix}
%\begin{pmatrix} A & 0 \vspace{0.5mm}\\
%0 & B \vspace{0.5mm}\\
%\hspace{1mm}\displaystyle
%\frac{\partial  f}{\partial i}(i^*, v^*) & 
%\displaystyle
%\frac{\partial  f}{\partial v}(i^*, v^*)\hspace{1mm}
%\end{pmatrix}
%\end{equation}
% remember $Q = -\partial f \partial i$
is nonsingular, a condition which by the implicit function theorem %readily
implies that this solution is locally unique. 
%In this context,
%some of the results of \cite{homoglin} can be easily extended to the nonlinear setting as follows.

\begin{theor} \label{th-main}
%Consider an uncoupled electrical circuit described by 
Let the model (\ref{reseq}) describe an uncoupled circuit.
The nondegeneracy of a given solution $(i^*, v^*)$ of these equations
%an uncoupled circuit described by the
%model (\ref{reseq})
is characterized
by the non-vanishing of the sum
  \begin{eqnarray} \label{kir}
\sum_{T \in {\cal T}}\left(\hspace{0.5mm} \prod_{j \in T }\frac{\partial f_j}{\partial v_j}(i_j^*, v_j^*) \prod_{k \in \overline{T} }\frac{\partial f_k}{\partial i_k}(i_k^*, v_k^*)\right),
    \end{eqnarray}
for any choice of the defining submersions $f_j(i_j, v_j)$, $j=1, \ldots, m$.
\end{theor}

\vs

\noindent {\bf Proof.} Choose first a fixed set of submersions $f_j(i_j, v_j)$
defining the %set of 
devices. 
The absence of coupling effects makes the matrices
of partial derivatives in (\ref{matrix0})
diagonal, and this gives the whole matrix (\ref{matrix0})
the form arising in (\ref{kir-hom}),
with $$P=\ \displaystyle \frac{\partial  f}{\partial v}(i^*, v^*), 
\ Q=-\displaystyle\frac{\partial  f}{\partial i} (i^*, v^*).$$
We may then apply Theorem \ref{th-pwmtt} above
%apply Lemma 1 and Theorem 1 from  \cite{homoglin}
to show that
%up to a nonzero constant $k_{AB}$, 
the determinant of (\ref{matrix0})
reads as %equals %the %so-called 
%multihomogeneous Kirchhoff (or tree-enumerator) polynomial 
  \begin{eqnarray} \label{kirbis}
k_{AB}\sum_{T \in {\cal T}}\left(\hspace{0.5mm} \prod_{j \in T }\frac{\partial f_j}{\partial v_j}(i_j^*, v_j^*) \prod_{k \in \overline{T} }\left(-\frac{\partial f_k}{\partial i_k}(i_k^*, v_k^*)\right)\right),
    \end{eqnarray}
for a certain nonzero constant $k_{AB}$.
%\begin{eqnarray} \label{kir2}
%K(p,q)=\sum_{T \in {\cal T}} %\left
%\big(\hspace{0.5mm} \prod_{j \in T }p_j \prod_{k \in \overline{T} } q_k \big).
%\right),
%\end{eqnarray}
The fact that all spanning trees define $m-n+1$ chords
%cotree branches 
implies that
(\ref{kirbis}) equals the expression depicted in 
(\ref{kir}) except for a nonzero factor $(-1)^{m-n+1}k_{AB}$,
%By choosing the homogeneous parameters $p_j$ and $q_k$ as the partial derivatives
%$\partial_{v_j}f_j$ and $\partial_{i_k}f_k$ we readily get (\ref{kir}) 
and the claim holds for the chosen set of submersions $f_j(i_j, v_j)$.

Additionally, from Theorem \ref{th-associates} and elementary properties of determinants,
%the choice of 
any other set of submersions $f_j(i_j, v_j)$ yields
non-vanishing factors $\gamma_j(i_j^*, v_j^*)$ in the determinant of (\ref{matrix0}).
%Including also the aforementioned constants, %$k_{AB}$, 
The resulting product defines
a nonzero factor $\gamma(i^*, v^*)$ which does not affect the nonsingular nature of (\ref{matrix0}).
\hfill $\Box$

\vspace{2mm}

\noindent The key contribution is that our formalism
%, including the homogeneous definition of the incremental resistance, %in homogeneous terms, 
allows one 
to handle %the function 
(\ref{kir})
in multihomogeneous form, yielding a %completely 
general result which applies to devices displaying
% accommodates
fully-implicit characteristics; this way
we %and %hence 
avoid 
any assumption on controlling
variables, which are thereby proved unnecessarily restrictive. 
%Certainly, by restricting the form of these characteristics to an explicit one, 
Of course, under additional hypotheses we get (as dehomogenizations)
known results in restricted scenarios.
%the claim of Theorem \ref{th-main} amounts to essentially known results in circuit theory. 
For instance, on voltage-controlled regions, 
where all derivatives $\partial f_j / \partial i_j$ are non-null (thus
allowing for local voltage-controlled descriptions $i_j=g_j(v_j)$ of
the devices),
%meaning that 
%all devices are ,
%(that is, if 
%(and, in particular, the circuit only has current sources), then $f_j(i_j, v_j)$ (cambiar g a f everywhere,
%keep g for conductance) can be written as $g_j(v_j)-i_j$ and (\ref{kir})
by multiplying (\ref{kir}) by the product of all inverses $(\partial f_j / \partial i_j)^{-1}(i^*, v^*)$, 
$j=1, \ldots, m$ or, in other terms,
by dehomogenizing %the function in 
(\ref{kir}),
we get, up to a $(-1)^{n-1}$ factor, 
%check
%$(-1)^mk_{AB}$ factor, NO
%amounts to 
the sum of the products of incremental conductances over the circuit spanning
trees. In this setting the results of subsection \ref{discussion} apply, including
the classical form of the matrix-tree theorem:
but we emphasize the fact that this is only one of the many possible dehomogenizations
which can be derived from the general expression (\ref{kir}).
%, this is consistent with 
%the well-known determinantal expansion of the nodal matrix \cite{chen} arising from the nodal equations
%$A g(A^{\tra} e) = 0$, %- A_s I_s $$, 
%where $e$ %and $I_s$ 
%stands for the node potentials (note that $g$ includes in particular the
%contribution of sources, which by the voltage-control requirement are implicitly assumed
%to be current sources).
% and the vector of currents injected by the sources.
%Indeed, by differentiating the left-hand side of the nodal equations one gets the weighted Laplacian
%matrix $A g'(A^Te)A^T$ and the classical form of the weighted matrix-tree theorem naturally applies.

\subsection{Dynamic circuits: the characteristic polynomial in multihomogeneous form}
\label{subsec-dynamic}

The ideas above are also useful in the context of dynamics circuits, involving reactive
devices and/or memristors. In the presence of 
capacitors and inductors (memristors are considered
in subsection \ref{subsec-memristors} below),
which for simplicity are assumed to be voltage- and current-controlled, respectively, the circuit
equations now read as
\begin{subequations}\label{dyneq}
\begin{eqnarray} 
C(v_c)v_c' & = & i_c \\
L(i_l)i_l' & = & v_l \\
0 & = & A_c i_c + A_l i_l + A_r i_r \label{dyneqc}\\
0 & = & B_c v_c + B_l v_l + B_r v_r \label{dyneqd}\\
0 & = & f(i_r, v_r).\label{dyneqe}
\end{eqnarray}
\end{subequations}
We keep a fully implicit formalism for resistors,
%In the latter equation, we are 
choosing again arbitrary submersions $f_j$ for their description in (\ref{dyneqe}).
Note that, as in the resistive context, the latter equation also accommodates independent
sources.
%of resistors and sources.
It is also worth remarking that when writing Kirchhoff equations in (\ref{dyneqc}) and
(\ref{dyneqd})
we have split the cut and cycle matrices by columns: e.g.\ $A_c$ denotes the submatrix
of $A$ defined by the columns which correspond to capacitors. Analogously, the vectors of 
currents and voltages are split according to the nature of the different devices. The context
should avoid any misunderstanding with subindices 
specifying a single branch (we use
systematically $j$ and $k$ for such subindices in what follows).
We assume, as before, that the circuit displays no coupling effects.

%Every j indicates a branch within the whole circuit

%**Careful notation $f$, ok if resistive branches are the first ones**

Focus the attention on an equilibrium point of (\ref{dyneq}),
defined by the conditions $i_c=0$, $v_l=0$, together with the 
corresponding restrictions 
%displayed in 
stemming from (\ref{dyneqc})-(\ref{dyneqe}). 
The spectrum of the linearization of the 
circuit equations at a given equilibrium $(i^*, v^*)$ 
is given by the
matrix pencil \cite{lmtbook, wsbook}
\begin{eqnarray} \label{pencil}
%\det \left(
\lambda \begin{pmatrix}
C(v_c^*)\hspace{-2mm}\vspace{0.99mm}& 0 & 0 & 0 & 0 & 0 \\
0 & L(i_l^*)\vspace{0.99mm} & 0 & 0 & 0 & 0\\ 
0 & 0 & 0 & 0\vspace{0.99mm} & 0 & 0\\ 
0 & 0 & 0 & 0 & 0\vspace{0.99mm} & 0\\ 
0 & 0 & 0 & 0 & 0 & 0
\end{pmatrix}
-
\begin{pmatrix}
0 & 0 & I_c & 0 & 0 & 0 \\
0 & 0 & 0 & I_l & 0 & 0\\ 
0 & A_l & A_c & 0 & A_r & 0\\ 
B_c & 0 & 0 & B_l & 0 & B_r\\ 
0 & 0 & 0 & 0 & \dsp\frac{\partial f}{\partial i_r}(i^*, v^*) & \dsp\frac{\partial f}{\partial v_r}(i^*, v^*)
%0 & 0 & 0 & 0 & -Q & P
\end{pmatrix},
%\right),
\end{eqnarray}
whose determinant
defines the characteristic polynomial of the circuit model (\ref{dyneq}) at equilibrium.
After elementary rearrangements, this determinant can be recast as
%\begin{eqnarray}
%\begin{pmatrix}
%0 & A_l & A_c & 0 & A_r & 0\\ 
%B_c & 0 & 0 & B_l & 0 & B_r\\ 
%\lambda C & 0 & -I_c & 0 & 0 & 0 \\
%0 & \lambda L & 0 & -I_l & 0 & 0\\ 
%0 & 0 & 0 & 0 & -Q & P
%\end{pmatrix} 
%\end{eqnarray}
%\begin{eqnarray}
%\det \begin{pmatrix}
%A_c & A_l & A_r &   0 & 0 & 0\\ 
%0 & 0 & 0 & B_c &   B_l  & B_r\\ 
%-I_c & 0 & 0 & \lambda C   & 0  & 0 \\
%0 & - \lambda L & 0 & 0 & I_l  & 0\\ 
%0 & 0 & -Q & 0 & 0  & P
%\end{pmatrix}, 
%\end{eqnarray}
\begin{eqnarray}
\det \begin{pmatrix}
A_c & A_l & A_r &   0 & 0 & 0\\ 
0 & 0 & 0 & B_c &   B_l  & B_r\\ 
-I_c & 0 & 0 & \lambda C(v_c^*)  & 0  & 0 \\
0 & - \lambda L(i_l^*) & 0 & 0 & I_l  & 0\\ 
0 & 0 & \dsp\frac{\partial f}{\partial i_r}(i^*, v^*) & 0 & 0  & \dsp\frac{\partial f}{\partial v_r}(i^*, v^*)
\end{pmatrix}, 
\end{eqnarray}
maybe up to a sign. %(depending on the parity of the column and row permutations).
Theorem \ref{th-pwmtt} then applies in a straightforward manner
to yield the result stated in Theorem \ref{th-charpol} below.
Within this result, for a given tree $T$ we denote by $T_c$, $T_l$ and $T_r$ 
(resp.\ $\overline{T}_c$, $\overline{T}_l$ and $\overline{T}_r$)
the sets of
%sub
indices of twig (resp.\ chord) %tree (resp.\ cotree)
capacitors, inductors and resistors, respectively.

%Given a tree $T$, we denote by $C(T)$, $L(T)$ and $R(T)$ 
%(resp.\ $C(\overline{T})$, $L(\overline{T})$ and $R(\overline{T})$) 
% which belong to the tree (resp.\ cotree).

\begin{theor} \label{th-charpol}
%Assume that a given $(i^*, v^*)$ is an equilibrium of (\ref{dyneq}). 
Up to a non-vanishing factor, the characteristic polynomial 
of the linearization of the circuit equations (\ref{dyneq})
at a given equilibrium reads as
\begin{equation} \label{CharPol}
\sum_{T \in {\cal T}} \prod_{\stackrel{j \in T_r }{\text{\tiny{$k\hspace{-0.6mm}\in\hspace{-0.6mm}\overline{T}_r$}}}} 
\frac{\partial f_j}{\partial v_j}(i_j^*,v_j^*) 
\left(-\frac{\partial f_k}{\partial i_k}(i_k^*,v_k^*)\right) 
%\prod_{k \in \overline{T}_r} Q_k 
\prod_{j \in T_c}(\lambda C_j(v_j^*))
\prod_{k \in \overline{T}_l}(\lambda L_k(i_k^*)).
\end{equation}
\end{theor}

%Maybe $\prod_j... \prod_k...$ also for resistors

\noindent Here we join the contribution of all resistors (either twigs
or chords) in a single product for notational simplicity;
note also that we are implicitly assuming that the circuit branches
are ordered in a way such that the indices
of resistors are the first ones, to make the subindex notation consistent
with the components of the map $f$.
This expression displays again a multihomogeneous nature in the 
parameters defining the incremental
resistances, that is, in the partial derivatives $\partial f_j / \partial v_j$ and
$\partial f_j/\partial i_j$. In other words, the choice of another defining submersion (say
for the $j$-th resistor) yields a non-vanishing common factor $\gamma_j(i_j^*, v_j^*)$ in the
expression above, since every branch contributes exactly one of such partial derivatives to each
monomial. Of course, this is the same idea that supports the homogeneous definition of the incremental
resistance in \ref{subsec-homogres}. Altogether, another set of submersions for the whole set
of resistors yields
a nonzero coefficient $\gamma(i_r^*, v_r^*)$ in the expression above.

For the sake of completeness it is useful to particularize the expression above to linear
problems. 
Provided that $j$ and $k$ are indices of a capacitive and an inductive branch, respectively,
we write as $C_j$, $L_k$ the corresponding scalar values of the 
%incremental NO, LINEAR
capacitance and %incremental IDEM 
inductance;
% at the $j$-th capacitive and the $k$-th inductive branches, respectively,
%evaluated at equilibrium; 
analogously, if the subindex $j$ corresponds to a resistive branch, we replace 
the partial derivatives $\partial f_j / \partial
v_j$ and $-\partial f_j / \partial i_j$ %above 
by 
%homogeneous 
the 
parameters $P_j$ and $Q_j$. %(mind the fact that $Q_j$ captures the $-$ sign).
%at equilibrium. 
%Diagonal matrices $C$, $L$, $P$, $Q$ comprising such entries...
%ALUSION A HOMOG NONLINEAR RESISTANCE
%Note that this notation
%is also meaningful in the linear context.
%
In this linear context, the characteristic polynomial then has the multihomogeneous form
\begin{equation} \label{CharPolLin}
\sum_{T \in {\cal T}} \prod_{\stackrel{j \in T_r }{\text{\tiny{$k\hspace{-0.6mm}\in\hspace{-0.6mm}\overline{T}_r$}}}} P_j Q_k
%\prod_{k \in \overline{T}_r} Q_k 
\prod_{j \in T_c}(\lambda C_j)
\prod_{k \in \overline{T}_l}(\lambda L_k).
\end{equation}
By dehomogenization we obtain Bryant's original expression for the characteristic polynomial,
derived in the seminal paper
\cite{bryant59}. The author assumes there that the circuit is linear and that 
all resistors are  current-controlled, being hence
described by the classical resistance parameters $R_j$. In our
framework,
this particular context
corresponds to the affine patch defined by the conditions $P_j \neq 0$ for all resistors
(cf.\ \cite{homoglin});
in this so-called current-controlled 
patch the resistances $R_j$ are well-defined as the quotients $Q_j/P_j$.
We then get Bryant's expression, namely,
\begin{equation}
\sum_{T \in {\cal T}} \prod_{k \in \overline{T}_r} R_k 
\prod_{j \in T_c}(\lambda C_j)
\prod_{k \in \overline{T}_l}(\lambda L_k)
\end{equation}
simply by dividing (\ref{CharPolLin}) by 
%$\prod_{j \in E_r}P_j$ 
%(here $E_r$: subindices corresponding to resistors, to avoid $j=1\ldots m_r$. Or assume, for notational
%simplicity, that the set of digraph branches are ordered so that the resistive branches are the first
%$m_r$ ones... and then use $j=1\ldots m_r$). OR 
$\prod_{j}P_j$, the product ranging
over all resistors. Note that, in each monomial,
the contribution of twig resistances disappears, and only the resistances of 
chord resistors matter.
It is a simple exercise to check that in the dual case (that is, in the voltage-controlled affine patch), the
corresponding expression for the characteristic polynomial is 
\begin{equation} 
\sum_{T \in {\cal T}} \prod_{j \in T_r } G_j
\prod_{j \in T_c}(\lambda C_j)
\prod_{k \in \overline{T}_l}(\lambda L_k),
\end{equation}
obtained after dividing (\ref{CharPolLin}) by the full
product $\prod_j Q_j$.
The reader may find 
the example of subsection \ref{subsec-finalexample} useful at this point; we refer him/her in particular
to the multihomogeneous expression (\ref{finalexample-hom}) for the characteristic
polynomial of the circuit analyzed there, and also to the voltage-controlled (conductance) dehomogenization
(\ref{finalexample-Gs}). We remark, finally, that all hybrid expressions, combining current- and voltage-controlled descriptions for different resistors,
can be derived from (\ref{CharPolLin}) in an analogous manner.

%Other notational options:
%\begin{equation}
%\sum_{T \in {\cal T}} \prod_{j \in T_r} P_j \prod_{j \in T_c}(\lambda C_j)
%\prod_{k \in \overline{T}_r} Q_k \prod_{k \in \overline{T}_l}(\lambda L_k)
%\end{equation}
%\begin{equation}
%\sum_{T \in {\cal T}} \prod_{j \in T_r} P_j 
%\prod_{k \in \overline{T}_r} Q_k 
%\prod_{j \in T_c}(\lambda C_j)
%\prod_{k \in \overline{T}_l}(\lambda L_k)
%\end{equation}

%\subsection*{Somewhere tema sources}

\subsection{Memristors}
\label{subsec-memristors}

Memristors and memristive devices
have attracted a lot of attention in Electronics
since the publication of the paper \cite{strukov} in 2008.
The existence of this device was predicted by Leon Chua in 
the seminal work \cite{chuamemristor71},
and the reader is referred to 
\cite{corinto2016, diventra09, itohchua08, itohchua2016, messias10, muthus10, muthuschua10, 
ishaq2017, memactive, tbwp, tetzlaffBook, vai} and references therein for background on this topic.

\subsubsection{Memristance and its homogeneous form}

Memristors
are defined by an intrinsically nonlinear flux-charge characteristic, which
is typically written either in the charge-controlled form
\begin{equation}\label{charcontr}
\varphi=h(\sigma)
\end{equation}
or in the flux-controlled one
\begin{equation}\label{fluxcontr}
\sigma=g(\varphi).
\end{equation}
Provided that the involved functions are differentiable,
in the former setting the {\em memristance}
is defined as $M(\sigma)=h'(\sigma)$, whereas in the second 
the {\em memductance} is $W(\varphi)=g'(\varphi)$. 
When both descriptions (\ref{charcontr}) and (\ref{fluxcontr}) are well-defined, the identities
$M(\sigma)=(W(h(\sigma)))^{-1}$ and $W(\varphi)=(M(g(\varphi)))^{-1}$ do hold.
In either case, 
we emphasize the nonlinear nature of the device; note, indeed, that
if $h$ or $g$ above are linear functions, then the memristance (resp.\ the 
memductance) is constant and the device simply behaves as a linear
resistor.

In this subsection we extend our previous results by using 
an implicit formalism for memristors: that
is, we avoid imposing any specific control variable
 by considering a fully-implicit
characteristic
\begin{equation}\label{full}
f(\sigma, \varphi)=0,
\end{equation}
where $f$ is a submersion 
which can be assumed to be globally defined on some open subset of $\R^2$,
consistently with the framework of Section \ref{sec-equiv}.
As before, this avoids restricting  in advance
the scope of the analysis and, again, 
when needed and by dehomogenization we can derive
results holding for explicit formulations such as (\ref{charcontr})
or (\ref{fluxcontr}). In this context, akin to the resistive context we 
define the {\em homogeneous memristance} at any point of the characteristic
as the pair of homogeneous coordinates
\begin{equation}
\left( \frac{\partial f}{\partial \varphi}(\sigma, \varphi):
-\frac{\partial f}{\partial \sigma}(\sigma, \varphi)\right),
\end{equation}
which (as a homogeneous pair)
is easily proved independent of the choice of the submersion $f$. Akin
to the resistive setting, at points where the partial derivative
$\partial f / \partial \varphi$ does not vanish, via the implicit
function theorem we may guarantee the existence of a local
description of the form (\ref{charcontr}), with memristance
$$M(\sigma)=-
\left(\frac{\partial f}{\partial \varphi}(\sigma, h(\sigma))\right)^{-1}
\frac{\partial f}{\partial \sigma}(\sigma, h(\sigma)).$$
Analogous remarks apply in the dual case.

\subsubsection{The characteristic polynomial of memristive circuits at equilibria}

The expressions defining the coefficients of the characteristic polynomial
in subsection \ref{subsec-dynamic} 
%framework of previous sections
can be naturally extended to the memristive context. We do that
in what follows, omitting some details for the sake of brevity.

Under an implicit description
of both resistors and memristors,
the circuit equations now take the form
\begin{subequations}\label{memeq}
\begin{eqnarray} 
C(v_c)v_c' & = & i_c \\
L(i_l)i_l' & = & v_l \\
\sigma_m' & = & i_m \\
\varphi_m' & = & v_m \\
0 & = & A_c i_c + A_l i_l + A_m i_m + A_r i_r \label{memeqe}\\
0 & = & B_c v_c + B_l v_l + B_mv_m + B_r v_r \label{memeqf}\\
0 & = & f_m(\sigma_m, \varphi_m)\label{memeqg} \\
0 & = & f_r(i_r, v_r).\label{memeqh}
\end{eqnarray}
\end{subequations}
%where $\sigma$ and $\varphi$ denote charge and flux, and the subindex $m$ corresponds to memristors.
%MAYBE subindex $w$, see below $m_w$... (number of memristors)
Equilibrium points are obtained after annihilating the right-hand
side; it is known in the circuit-theoretic literature that such equilibria
are never isolated in the presence of at least one memristor
\cite{corinto2016, messias10, tbwp}.

Now, using a Schur reduction \cite{hor0}, the characteristic polynomial at a given equilibrium may in this context be checked to be defined by 
\begin{eqnarray}
\lambda^{m_w}\det \begin{pmatrix}
A_c & A_l & A_m & A_r & 0 &  0 & 0 & 0\\ 
0 & 0 & 0 & 0 & B_c &   B_l & B_m & B_r\\ 
-I_c & 0 & 0 & 0 & \lambda C   & 0  & 0 & 0 \\
0 & - \lambda L & 0 & 0  & 0 & I_l  & 0 & 0 \\ 
0 & 0 & -Q_m & 0 & 0 & 0 & P_m & 0 \\
0 & 0 & 0 & -Q_r & 0 & 0 & 0 & P_r
\end{pmatrix} 
\end{eqnarray}
where $m_w$ is the number of memristors.
For notational simplicity, we are writing 
%(**maybe add double subindex, j**)
\begin{equation} \label{PmQm}
P_m = \frac{\partial f_m}{\partial \varphi_m}, \ Q_m=
-\frac{\partial f_m}{\partial \sigma_m}
\end{equation}
for memristors, and
\begin{equation} \label{PrQr}
P_r = \frac{\partial f_r}{\partial v_r}, \ Q_r=
-\frac{\partial f_r}{\partial i_r}
\end{equation}
for resistors;
the dependence on the 
different circuit variables (charge, flux, voltage and current) is
also omitted for simplicity. As in previous sections we are
excluding coupling effects within each device type and therefore all
the matrices in (\ref{PmQm}) and (\ref{PrQr}) are diagonal.

Proceeding as above, the characteristic polynomial now has the expression (always up to a 
non-vanishing factor)
\begin{equation} \label{memPolChar}
\lambda^{m_w}\sum_{T \in {\cal T}} \prod_{\stackrel{j \in T_r \cup T_m}{\text{\tiny{$k\hspace{-0.6mm}\in\hspace{-0.8mm}\overline{T}_r
\hspace{-0.6mm}\cup\hspace{-0.5mm}\overline{T}_m$}}}} \hspace{-1.5mm} P_j Q_k
%\prod_{k \in \overline{T}_r} Q_k 
\prod_{j \in T_c}(\lambda C_j)
\prod_{k \in \overline{T}_l}(\lambda L_k)
\end{equation}
%$\sigma$ and $\varphi$
where we denote by $P_j$ and $Q_j$ the 
individual homogeneous parameters of both resistors and memristors.
Note incidentally that the $P_jQ_k$ products range in (\ref{memPolChar})
 over the full set of resistors and memristors which, again,
provide a homogeneous contribution to every monomial, 
the appearance of either $P$ or $Q$ depending
on the actual location of the device in the corresponding tree or cotree.
Conventional expressions in terms of the 
individual memristances and/or memductances can be derived as in subsection
\ref{subsec-dynamic}
by setting
$M_j=Q_j/P_j$ or $W_j=P_j/Q_j$ in the patches
where the respective denominators do not vanish (an example
is discussed below).
%Maybe also formula with partial derivatives
Finally, worth remarking is also 
the presence of a $\lambda^{m_w}$ term %(we use $m_w$ for the total
%number of memristors), 
which is responsible for a zero eigenvalue
whose algebraic multiplicity equals the number of memristors
(find details  in this regard in \cite{tbwp}).
%,  consistently with the aforementioned fact that equilibria are not %never 
%isolated.
%in memristive circuits.
%a result which is already
%well-known in the literature (see e.g.\ 
%\cite{messias10, tbwp}.

\subsubsection{Example}
\label{subsubsec-exmem}

%CHAR POL WITH DERIVATIVES, NOW LINEAR MAKES NO SENSE (memristors)

%We illustrate the form that this polynomial takes this practice by looking at a small-scale
%example, namely the circuit displayed in Figure **

\begin{figure}[htb] \label{fig-ishaq}
%\parbox{5in}{%
%\hspace{82mm}
\begin{center}
\epsfig{figure=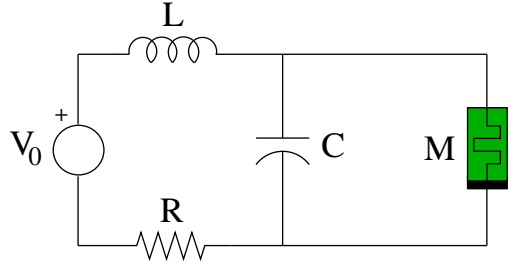, width=0.4\textwidth}%, angle=270}
%}
%\vspace{-3mm}
\end{center}
\caption{Murali-Lakshmanan-Chua circuit.\hspace{-2mm}}
\label{fig-MLC}
\end{figure}

\noindent We illustrate the form of this 
polynomial in terms of the example depicted in Figure 1, taken from
\cite{ishaq2017}. 
Specifically, our purpose is to detail in an example 
how the different coefficients in the multihomogeneous characteristic polynomial 
(\ref{memPolChar})
%in  form
can be examined in terms of the spanning-tree structure of the circuit,
and show the way in which eventual restrictions in the controlling variables
are captured via dehomogenization. First,
it is easy to check 
that only spanning trees including the voltage source define nonzero 
products in (\ref{memPolChar}):
%(since a voltage source in the cotree contributes a zero factor to the corresponding
%product because of the condition $\partial f/\partial i=0$ in its defining function **Nmg**):
the attention will be henceforth restricted to trees including 
the voltage source without further explicit mention.
Additionally, the absence of loops defined exclusively by voltage sources and capacitors, and of cutsets
just defined by current sources (absent in this example)
and inductors, guarantees that, generically, the order
of the circuit (that is, the degree of the characteristic polynomial) equals the number
of memristors and reactive elements (see \cite{memactive}). 
Actually, the leading
term of the characteristic polynomial is defined by the set of spanning trees including 
the capacitor and excluding the inductor. There is only one
such tree, depicted on the left of Figure 2. Since the resistor is in the tree but the
memristor is not, this yields a leading coefficient of the form
$$LCP_rQ_m\lambda^3.$$
According to (\ref{memPolChar}) the coefficient of $\lambda^2$ is defined
in this case by the spanning trees which include either both or none of the reactive elements.
These are the remaining trees in Figure 2, and yield the term
$$(CQ_rQ_m + LP_rP_m)\lambda^2.$$
Finally, the coefficient of $\lambda$ is defined by the trees which include the inductor but
not the capacitor. These are displayed in Figure 3. The corresponding term is
$$(P_rQ_m + P_mQ_r)\lambda$$
and the full characteristic polynomial reads as
\begin{equation}\label{polCharEx}
LCP_rQ_m\lambda^3 + (CQ_rQ_m + LP_rP_m)\lambda^2 + (P_rQ_m + P_mQ_r)\lambda.
\end{equation}
\begin{figure}[htb]
%\parbox{5in}{%
%\hspace{82mm}
\begin{center}
\hspace{-10mm}\parbox{0.33\textwidth}{\epsfig{figure=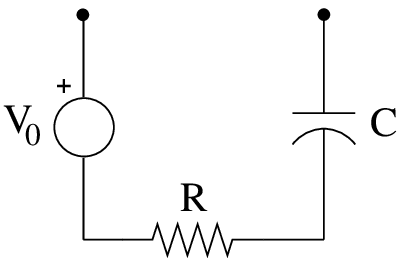, width=0.275\textwidth}}%, angle=270}
\hspace{-3mm}\parbox{0.33\textwidth}{\vspace{-7.5mm}\epsfig{figure=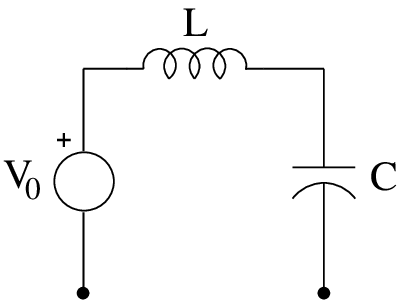, width=0.275\textwidth}}%, angle=270}
\hspace{-3mm}\parbox{0.33\textwidth}{\epsfig{figure=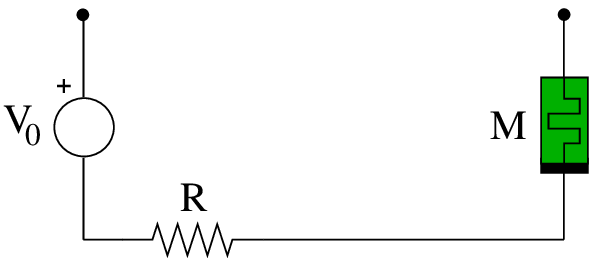, width=0.4\textwidth}}%, angle=270}%}
%\vspace{-3mm}
\end{center}
\caption{Spanning trees yielding the $\lambda^3$ and $\lambda^2$ terms.\hspace{-8mm}}
%\label{fig-MLC}
\end{figure}

\begin{figure}[htb]
%\parbox{5in}{%
%\hspace{82mm}
\begin{center}
\parbox{0.33\textwidth}{\epsfig{figure=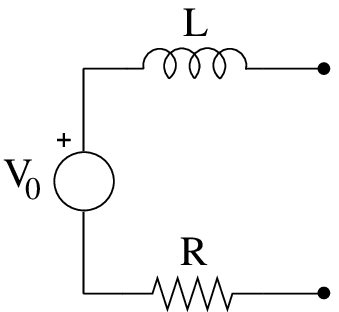, width=0.225\textwidth}}%, angle=270}
\parbox{0.33\textwidth}{\vspace{-1mm}\epsfig{figure=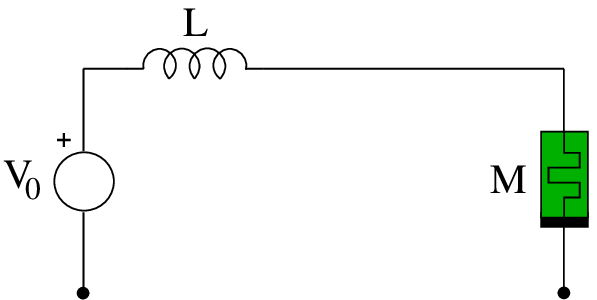, width=0.4\textwidth}}%, angle=270}
%}
%\vspace{-3mm}
\end{center}
\caption{Spanning trees defining the term in $\lambda$.\hspace{-18mm}}
%\label{fig-MLC}
\end{figure}

We emphasize the fact that our approach applies without the need
to assume neither a specific voltage- or current-controlled form for resistors 
nor a flux- or charge-controlled description of memristors; this means that (\ref{polCharEx})
holds without any such restrictions.
Again, particular forms can be obtained
by dehomogenization under specific hypotheses: for example, in \cite{ishaq2017} the authors assume
that the resistor is linear and current-controlled and that the memristor is flux-controlled. 
This case is obtained by dividing (\ref{polCharEx}) by $P_r$ and $Q_m$ to get the resistance $R$ 
as the quotient $Q_r/P_r$ and the memductance $W$ as $P_m/Q_m$. This transforms (\ref{polCharEx}) into
\begin{equation}\label{polCharExBis}
LC\lambda^3 + (RC + LW)\lambda^2 + (1 + RW)\lambda,
\end{equation}
an expression which is obtained (with normalized parameters) in \cite{ishaq2017}
only after linearizing a state-space model of the
circuit. Even if such assumptions are reasonable in many practical cases, notice that
some key information is lost when turning from (\ref{polCharEx}) to (\ref{polCharExBis}).
In particular, (\ref{polCharExBis}) provides no information about the dual 
setting to the one considered in \cite{ishaq2017}. Indeed,
should the resistor be voltage-controlled (and described by the conductance $G$)
and/or the memristor charge-controlled (with memristance $M$),
singularities due to the eventual vanishing of 
$G$ or $M$ would not be captured in (\ref{polCharExBis}). This means that
there is no chance to describe such singularities (and the corresponding order
reduction, possibly responsible for impasse phenomena: see \cite{corinto2016}
in this regard) by looking at (\ref{polCharExBis}). By
contrast, the general form depicted in 
(\ref{polCharEx}) captures these cases in a smooth manner
simply by fixing respectively the parameter
values $P_r=0$ or $Q_m=0$, both of which annihilate the leading term of (\ref{polCharEx}).
%; recall that $G=P_r/Q_r$ and $M=Q_m/P_m$ in regions
%where the denominators do not vanish.

\section{
 %5. Application: s
  Simple stationary bifurcation points in nonlinear circuits}
\label{sec-bif}

The formalism introduced in previous sections provides 
a set of useful tools for qualitative studies of the dynamics of
 nonlinear circuits. In particular, in what follows
we apply our framework to a bifurcation analysis, focusing
%The focus will be 
on %so-called {\em 
steady bifurcations
%\cite{seydel} 
of dynamic circuits without memristors. The goal is to provide
a characterization of so-called simple stationary bifurcation
points in graph-theoretic terms, that is, a characterization
explicitly formulated in terms of
certain  assumptions on the topology of the circuit
and the electrical properties of the devices. This
extends %(with the tools just introduced)
previous studies of other bifurcations in the circuit context \cite{saddlenode, tbwp}.
%Specifically, we will make systematic use of the 
%tree-based, homogeneous determinantal formulae yielding
%the form (\ref{CharPol}) for the characteristic polynomial. 

\subsection{Stationary bifurcations}
\label{subsec-seydel}

%%5.1 Seydel conditions for a SSBP can be examined in tree-terms (may 2018,
%check top hypotheses may be relaxed w.r.t. TB). TB: additionally
%no PIEs + fold (but these cannot be formulated in terms of trees, avoid).

%P. 74 Seydel

Consider an explicit ODE $x'=F(x, \mu)$, $x \in \Omega \subseteq \R^n$, and 
let the origin (assumed to belong to $\Omega$) be an equilibrium for any
$\mu$ on a neighborhood of a given $\mu_0$. In this setting, the point $(0, \mu_0)$ is called a
%$(y_0, \lambda_0)$ is a 
{\em simple stationary bifurcation point} if the
following conditions hold:

%(1) Seydel b$f (y_0, \lambda_0) = 0$;

%(2) Seydel $rank f_y(y_0, \lambda_0) = n - 1$; -later in Chapter 5 implicitly uses simple zero eigenvalue
\begin{itemize}
\item[(i)] $\lambda=0$ is a simple eigenvalue of $F_x(0,\mu_0)$; and

%(3) $f_{\lambda}(y_0, \lambda_0) \in \im f_y(y_0, \lambda_0)$; %HOLDS TRIVIALLY and

%(4) exactly two branches of stationary solutions intersect with two distinct tangents. Guaranteed by the following, which means $b \neq 0$ in Seydel's chapter 5
\item[(ii)] $F_{x\mu}(0,\mu_0) v \notin \im F_x(0,\mu_0)$ for $v \in \ke F_x(0,\mu_0)-\{0\}$.
\end{itemize}
Here we are using the subscripts $x$ and $\mu$ to denote partial differentiation.
Find detailed introductions to stationary bifurcations in \cite{perko, seydel}: note only that the
conditions above can be stated in this straightforward form because of the assumption
that the origin is, at least locally, an equilibrium for any $\mu$.
For our present purposes, the important remark is that 
%Note that \cite{seydel} uses a slightly broader definition, cf.\ Definition 2.7 there; 
%however, under the 
%hypothesis that the origin is (locally) an equilibrium for any $\mu$, the conditions later considered
%in \cite{seydel}[5.***] can be checked to amount to the ones above (this is on my sheets). 
%It is also worth emphasizing that, 
under additional 
requirements 
(namely, a second order transversality condition and the absence of other eigenvalues in the imaginary axis),
a simple stationary bifurcation point as defined above is well-known to be responsible for a 
transcritical bifurcation \cite{perko}, 
displaying an exchange of stability between two equilibrium branches; this will
be the case in the example considered at the end of this section.

%Hypothesis (4) is unnecessarily strong, and preliminary. It aims at the ``two,''
%and will be replaced by a weaker criterion based on second-order derivatives
%in Section 5.5.2, see Theorem 5.7. Pages 226-229/230

\subsection{A circuit-theoretic characterization of the bifurcation}
\label{subsec-steady}

%Our present goal is to show how the homogeneous 
%formalism above and, in particular, determinantal formulae can be used in the topological characterization of certain qualitative properties of nonlinear
%circuits. 
Theorem \ref{th-bif} below presents a graph-theoretic characterization of 
the existence of simple stationary bifurcation points  in nonlinear circuits,
in the presence of
certain topological configurations.
The
bifurcation phenomenon
will be related to a distinguished resistor
(supposed w.l.o.g.\ to be the first one), which
is here assumed to be voltage-controlled, with
a characteristic of the form
\begin{equation}
i = g(v, \mu) = \mu v + \tilde{g}(v),\label{nonpassive}
\end{equation}
with $\tilde{g}(v)=o(|v|)$. Note that $\mu$ is the incremental conductance at the origin and 
our purpose is to characterize
the bifurcation conditions with this conductance behaving as the bifurcation parameter. 
Because of the form of (\ref{nonpassive}) we have $g(0,\mu)=0$. 
The remaining resistors are assumed to be characterized
by arbitrary submersions $f_j(i_j, v_j).$ 
%for $j=2, \ldots, m_r$ (cf.\ (\ref{dyneq}).
%** OJO TEMA SUBINDICES Tr ETC **

In the statement of Theorem \ref{th-bif} we split the hypotheses into those referring to
the circuit topology (hypotheses T1 and T2) and the ones involving
the electrical properties of the constitutive devices (D1, D2 and D3). We will make use of the notion
of a {\em proper tree}, which is a spanning tree including all capacitors and no inductors.
We denote the family of proper trees by ${\cal T}_p$.

%Recall:
%\begin{equation}
%\sum_{T \in {\cal T}} \prod_{\stackrel{j \in T_r }{\text{\tiny{$k\hspace{-0.6mm}\in\hspace{-0.6mm}\overline{T}_r$}}}} P_j Q_k
%%\prod_{k \in \overline{T}_r} Q_k 
%\prod_{j \in T_c}(\lambda C_j)
%\prod_{k \in \overline{T}_l}(\lambda L_k)
%\end{equation}

%\

\begin{theor} \label{th-bif}
Consider an uncoupled electrical circuit for which the following hypotheses hold: 
\begin{enumerate}
\item[\rm{T1.}] There are no loops or cutsets defined by only one type of reactive elements (capacitors or inductors).
\item[\rm{T2.}] The bifurcating resistor (governed by (\ref{nonpassive})) 
forms a cutset together with one or more capacitors.
\item[\rm{D1.}] The characteristics of all resistors meet the origin.
\item[\rm{D2.}] All circuit devices but the bifurcating resistor are 
strictly locally passive at the origin.
\item[\rm{D3.}] The sum 
$\dsp\sum_{T \in {\cal T}_p}\left(\hspace{0.5mm} \prod_{j \in T_r}\frac{\partial f_j}{\partial v_j}(0,0) \prod_{k \in \overline{T}_k}\frac{\partial f_k}{\partial i_k}(0, 0)\right)\hspace{-1mm},$
ranging over proper trees, does not vanish.
%The C-proper tree sum [] does not vanish at the origin, or: the origin is a regular point. %(ie the origin is not an impasse point).
%DOES NOT HOLD AUTOMATICALLY BECAUSE OF THE PASSIVITY ASS BECAUSE $\mu$ MIGHT ENTER ALL C-PROPER TREES
\end{enumerate}

\vspace{2mm}

\noindent Then $\mu_0=0$ yields a simple stationary bifurcation point at the origin.
\end{theor}

\vspace{2mm}

\noindent {\bf Proof.} 
Note first that because of the condition
$g(0,\mu)=0$ holding for the bifurcating resistor (cf.\ (\ref{nonpassive})), together with hypothesis
D1, the origin happens to be an equilibrium
for any value of the parameter $\mu$. We therefore need to check that conditions (i) and (ii)
in subsection \ref{subsec-seydel} hold, in the understanding that the vector field
$F$ arising there 
comes from the reduction of the differential-algebraic model (\ref{dyneq}), 
in the terms detailed later.

Let us first focus on condition (i), which in our context amounts to saying that the
characteristic polynomial (\ref{CharPol}) has a null independent term at the origin
(so that $\lambda=0$ is an eigenvalue) with the coefficient of the term in $\lambda$ not
vanishing (this making $\lambda=0$ a simple eigenvalue). To check this we have to examine the form
of such coefficients in light of the hypotheses above.

The independent term  in (\ref{CharPol}) is given by the sum %of products of the form
\begin{equation}\label{indepTerm}
%\left(
\sum_{T \in {\cal T}_p^*} \prod_{j \in T_r}\frac{\partial f_j}{\partial v_j}(0,0) \prod_{k \in \overline{T}_k}
\left(-\frac{\partial f_k}{\partial i_k}(0, 0)\right)\hspace{-0.5mm},
\end{equation}
ranging over spanning trees which include all inductors and no capacitors
(we denote the family of such trees by ${\cal T}_p^*$). At least one
such tree exists because of the absence of inductor loops and capacitor cutsets assumed in T1.
%corresponds to the sum of conductance products in L-proper trees, because all capacitors must be in the cotree and all inductors in the tree for $\lambda$ not
%to appear as a factor. At least one such tree  exists because of the assumed absence of L-loops and C-cutsets; moreover, it is not difficult to show that 
The key fact at this point is that
 all trees in ${\cal T}_p^*$ have
 the bifurcating resistor as a twig. This is a
consequence of an elementary 
graph-theoretic property, namely, that a spanning tree must include at least one branch
from every cutset; since capacitors are necessarily excluded from the spanning 
trees in ${\cal T}_p^*$, %yielding the independent term (\ref{indepTerm}), 
hypothesis T2 implies
that the bifurcating resistor enters all such trees. 
Now, by setting $f_1=-i_1+\mu v_1 + \tilde{g}(v_1)$ for the bifurcating
resistor (cf.\ (\ref{nonpassive})), 
we have $\partial f_1/\partial v_1=\mu$ 
at the origin, so that $\mu$ is a common factor to all the summands in
%sums of products of the form
(\ref{indepTerm}).
Additionally, by the strict passivity assumption stated in hypothesis D2, all
summands
%the remaining factors 
in (\ref{indepTerm}) must display
%yield 
the same sign, regardless of the actual choice of the defining submersions. %in each monomial. 
Indeed, the passivity condition implies that 
$\partial f_j/\partial v_j$ and $-\partial f_j/\partial i_j$ have at the origin
the same sign for each $j$
(exception made of $j=1$, that is, of the bifurcating resistor),
%in the sum, 
and the claim then follows easily from the multihomogeneous form
of the polynomial (\ref{kir-hom}).
%NO : notice, indeed, 
%that this sign is actually characterized by the number of resistors in each cotree, but
%this number is the same for
%all the spanning trees defining the products (\ref{indepTerm}).
This means that the independent term of the characteristic
polynomial simply reads as $\mu k$ for some non-null constant $k$. In particular, this independent
term vanishes at the bifurcation value $\mu=0$.

%the sum above
%reads
%$$\mu $$%\left(
%\prod_{j \in T_r, j \neq 1}\frac{\partial f_j}{\partial v_j}(0,0) \prod_{k \in \overline{T}_k}\frac{\partial f_k}{\partial i_k}(0, 0),$$

%*** this follows from the exclusion of twig capacitors together
%with the assumed existence of a cutset formed by this resistor and some capacitors (from elementary properties of a graph one can 
%see that ). Together with the fact that all remaining devices are strictly locally passive, this implies that the independent term
%reads as $\mu k$ for a positive constant $k$. In particular this means that at $\mu=0$ the value $\lambda=0$ is indeed a root of the polynomial and 
%we indeed have a zero eigenvalue. [TOO MUCH TEXT, MAYBE BRING POLYNOMIAL]

Regarding the term in $\lambda$ within the characteristic polynomial (\ref{CharPol}), 
%proceeding as above 
%from (\ref{CharPol}) it follows that the coefficient of $\lambda$ 
notice that this term comes from all spanning trees
which include either exactly one capacitor and all inductors in the tree, 
or exactly one inductor and all capacitors in the cotree. 
Again, one can
check that at least one of such spanning trees does exist, since
the replacement of the 
bifurcating resistor in any of the spanning trees in the paragraph above
(namely, the ones defining the independent term of the characteristic polynomial)
by one capacitor from the cutset defined in hypotheses T2
results in a spanning
tree which includes exactly one capacitor and all inductors. 
The contribution of this term to the coefficient is non-zero because of the absence of 
the bifurcating resistor from the tree
(so that it contributes a factor $-\partial f_1/\partial i_1=1$, with $f_1$ defined above)
together with the assumption that all remaining devices are strictly locally
passive. Since other spanning trees contribute to the sum terms which are zero (if such
trees include the bifurcating resistor) %as a twig)
or, as in the preceding paragraph, have otherwise the same sign as the one above (if they do not),
one concludes that the coefficient of the
term in $\lambda$ is non-null and this means that the 
zero eigenvalue is indeed a simple one.

It remains to check condition (ii) from subsection  \ref{subsec-seydel},
that is, $F_{x \mu} v \notin \im F_x$ for $v \in \ke F_x -\{0\}$ (with all derivatives
evaluated at the origin). From well-known properties of matrix analysis
(see \cite{hor0} or, specifically, Lemma 1 in \cite{tbwp}), this condition
can be equivalently assessed as
\begin{equation} \label{derivada}
\frac{\partial  \det F_x}{\partial \mu}(0, \mu_0) \neq 0.
\end{equation}
In our setting $F$ stands for the vector field defining an explicit local reduction of the differential-algebraic
model (\ref{dyneq}) at equilibrium: here we use hypothesis D3, which guarantees
that the leading coefficient of the characteristic polynomial (\ref{CharPol}) 
does not vanish; 
worth remarking is that the capacitances and inductances are positive at the origin because
of the strict local passivity hypothesis D2. This means that the matrix pencil (\ref{pencil}) has nilpotency
index one (cf.\ \cite{bre1, lmtbook}) and a state reduction 
of (\ref{dyneq}) (or, in other words, the vector field $F$ above) is thus well-defined
in terms of $v_c$ and $i_l$ 
\cite{wsbook}. 

For our purposes there is no need to compute explicitly this
reduced vector field; it is enough to use %its linearization
the Jacobian matrix $F_x$ at the origin, which by construction
is the Schur reduction \cite{hor0} $M_1/M_2$, with
\begin{eqnarray*} %\label{M1}
M_1 =
\begin{pmatrix}
0 & 0 & (C(0))^{-1} & 0 & 0 & 0 \\
0 & 0 & 0 & (L(0))^{-1} & 0 & 0\\ 
0 & A_l & A_c & 0 & A_r & 0\\ 
B_c & 0 & 0 & B_l & 0 & B_r\\ 
0 & 0 & 0 & 0 & -Q & P
\end{pmatrix}
\end{eqnarray*}
coming from the linearization of %the right-hand side of 
(\ref{dyneq}), and
\begin{eqnarray*} %\label{M2}
M_2 =
\begin{pmatrix}
A_c & 0 & A_r & 0\\ 
0 & B_l & 0 & B_r\\ 
0 & 0 & -Q & P
\end{pmatrix}.
\end{eqnarray*}
In both matrices we denote $P=\partial f_r/\partial v_r$, $Q=-\partial f_r/\partial i_r$, 
with the derivatives being always evaluated at equilibrium. We leave it to the
reader to check that $M_2$ is non-singular
because of hypothesis D3. Now, a well-known property of the Schur reduction says that
\begin{eqnarray} \label{Schurdet}
\det M_1 = \det M_2 \det F_x.
\end{eqnarray}
Additionally, since as shown above the matrix pencil (and hence the linearized
vector field $F_x$) has a zero eigenvalue at equilibrium, 
we have $\det F_x=0$ at the origin and,
by differentiating (\ref{Schurdet})
we may hence evaluate (\ref{derivada}) as
\begin{equation} \label{derivadabis}
\frac{\partial \det M_1}{\partial \mu} \neq 0.
\end{equation}
However, up to a non-zero factor the determinant of $M_1$ equals
\begin{eqnarray} \label{enesima}
  \det \begin{pmatrix}
A_c & A_l &  A_r & 0 & 0 & 0\\
0 & 0 & 0 & B_c & B_l & B_r  \\
I_c & 0 & 0 & 0 & 0 & 0 \\
0 & 0 & 0 & 0 & I_l & 0 \\
0 & 0 & -Q & 0 & 0 & P
\end{pmatrix}.
\end{eqnarray}
Using Theorem \ref{th-pwmtt} one can check that the latter determinant is given 
by the sum %of products of the form 
(\ref{indepTerm})
(because tree capacitors or cotree inductors
yield a null factor in the corresponding term, in light of the form of the matrix in (\ref{enesima})),
again possibly up to a sign.
As shown in the first part of the proof, this sum is
given by $\mu k$ for some non-vanishing constant $k$. This means
that the derivative in (\ref{derivadabis}) is not zero, as we aimed to show.
%equals $\pm k \neq 0$, 
From 
(\ref{derivadabis}) it follows that
(\ref{derivada}) is met: therefore, condition (ii) also holds and  the proof is complete.
%as we aimed to show.
%\vspace{2mm}
\hfill 
$\Box$

%Non-singular eq: nonsingularity of
%\begin{eqnarray} 
%\begin{pmatrix}
% A_l &  A_r & 0 & 0 \\
%0 & 0 & B_r & B_c \\
%0 & -Q_r & P_r & 0
%\end{pmatrix}
%\end{eqnarray}
%a Schur (up to signs) of
%\begin{eqnarray} 
%\begin{pmatrix}
%A_c & A_l &  A_r & 0 & 0 & 0\\
%0 & 0 & 0 & B_r & B_c & B_l \\
%0 & 0 & -Q_r & P_r & 0 & 0\\
%I_c & 0 & 0 & 0 & 0 & 0 \\
%0 & 0 & 0 & 0 & 0 & I_l 
%\end{pmatrix}
%\end{eqnarray}
%and this corresponds to 0's in the twig C's and in the link L's ok

%outline (6.8.18):

%- We work on an index one setting (away from impasse set) 

%- Equilibrium for any $\mu$ OK

%- Zero eigenvalue OK

%- Simple OK

%- $f_{\mu x} v \notin \im f_x$ (coefficient de $\mu$ in L-proper sum)  OK

%\hfill $\Box$

\subsection{Example} 
%RC tank: simultaneously SSBP + SIB 19.5.18... justifies formalism? NO, another option
%Here: 22.5.18, VdPol with a resistance in parallel to L
\label{subsec-finalexample}

Finally, we illustrate the result stated in Theorem \ref{th-bif} by means of an elementary example,
depicted in Figure 4. This circuit may also help the reader understand the 
expression (\ref{CharPol}) derived for %We use this circuit to show further how to compute 
the characteristic polynomial  in subsection \ref{subsec-dynamic}. 

\begin{figure}[htb] \label{fig-steady}
%\parbox{5in}{%
%\hspace{82mm}
\begin{center}
\epsfig{figure=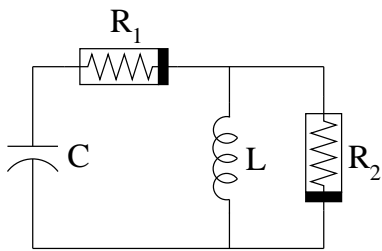, width=0.3\textwidth}%, angle=270}
%}
%\vspace{-3mm}
\end{center}
\caption{A circuit undergoing a transcritical bifurcation. \hspace{-2mm}}
%\label{fig-steady}
\end{figure}

\noindent The resistor labelled as $R_1$ will behave as the bifurcating device, and is therefore
assumed to be governed by an expression of the form depicted in (\ref{nonpassive}), that is,
$$i_1 = g_1(v_1, \mu)=\mu v_1 + \tilde{g}(v_1).$$ Note that it forms a cutset together
with the capacitor, since the removal of both branches disconnects the
circuit.
The key topological hypothesis in Theorem \ref{th-bif} (hypothesis
T2) is therefore met. For simplicity,
we take $\tilde{g}(v_1)=v_1^2$. For later use, let additionally $f_1(i_1, v_1, \mu)$ stand
for $i_1 - g_1(v_1, \mu)$. The capacitor and the inductor are assumed to be linear and
strictly passive, with positive capacitance and inductance $C$ and $L$, respectively. Finally,
we give %for the moment 
the second resistor an implicit form, that is, we
write its characteristic in the form $f_2(i_2, v_2)=0$, and assume $f_2(0, 0)=0$.

Defining directions by choosing top-down orientations in the branches
depicted in vertical in the figure, the circuit equations can be easily written as follows:
\begin{subequations} \label{exsteady}
\begin{eqnarray}
Cv_c' & = & g_1(v_1, \mu) \label{exsteadya}\\
L i_l' & = & v_1 + v_c \label{exsteadyb}\\
0 & = & i_l + i_2 + g_1(v_1, \mu) \label{exsteadyc}\\
0 & = & f_2(i_2, v_1+v_c),\label{exsteadyd}
\end{eqnarray}
\end{subequations}
where we have performed some elementary simplifications with respect to the
general model (\ref{dyneq}) (in particular,
the voltage across the inductor and the second resistor
is written in terms of $v_1$ and $v_c$ in (\ref{exsteadyb}) and (\ref{exsteadyd})).

In this setting, the circuit can be easily checked to have
two equilibrium loci, defined by $v_1 = 0$ and $v_1=-\mu$
(together with $v_c = -v_1,$ $i_l=i_2=0$), which intersect at the origin. 
%Using the map $f_1$ defined above and 
For the sake of simplicity in the notation, we introduce
the parameters
$$ P_j = \frac{\partial f_j}{\partial v_j}, \ Q_j = - \frac{\partial f_j}{\partial i_j}, \ j=1, 2,$$
with $f_1$ and $f_2$ defined above.
The multihomogeneous form (\ref{CharPol}) of the characteristic
polynomial reads at equilibrium as
\begin{equation}
\label{finalexample-hom}
LC (P_1Q_2 + Q_1 P_2) \lambda^2 + (CQ_1Q_2 + L P_1P_2) \lambda + P_1Q_2,
\end{equation}
an expression that the reader may derive either from linearizing (\ref{exsteady}) or by examining the 
structure of the spanning
trees of the circuit, as we did in subsection \ref{subsubsec-exmem}.

To illustrate more easily the notions used in Theorem \ref{th-bif},
let us restrict the attention to cases in which both resistors admit a voltage-controlled
description. This is equivalent to dehomogenize the polynomial above by dividing by $Q_1Q_2$, to get
\begin{equation}
\label{finalexample-Gs}
LC (G_1 + G_2) \lambda^2 + (C + L G_1G_2) \lambda + G_1.
\end{equation}
% (Further comment on $G_1$, even IFT quotient).
%Let us further focus the attention on the origin, where $G_1 = 0$.
The non-vanishing of the leading coefficient amounts
to the condition $LC(G_1+ G_2) \neq 0$: both $L$ and $C$ are non-zero
(actually positive)
by hypothesis, whereas the non-vanishing condition on $G_1+ G_2$ reflects in this case
the general requirement
expressed by  hypothesis D3 in Theorem
\ref{th-bif}. Note in this direction that the circuit has two proper trees, defined by the capacitor
and each one of the two resistors, which are responsible for the $G_1 + G_2$ expression above
(or, in greater generality, for the term
$P_1Q_2 + Q_1 P_2$ in the multihomogeneous expression (\ref{finalexample-hom})).
%FIGURE?. 
Focusing the attention on the bifurcation value $\mu=0$, we have
$G_1=0$ at the origin, and 
the requirement $G_1+ G_2 \neq 0$ is in this case actually
met by the strict passivity assumption on the second resistor
(which means that $G_2 > 0$). The reader may check that all the remaining hypotheses of Theorem
\ref{th-bif} are met and, therefore, a simple stationary bifurcation point is expected
to be displayed at the origin. %And even if this is not addressed in subsection \ref{subsec-steady} 
This is actually the case and, moreover,
a transcritical bifurcation occurs at that point: indeed,
the system experiences an exchange of stability, since
the equilibrium branch defined by $v_1 =0$ can be checked to be 
asymptotically stable for $\mu >0$ but unstable
for $\mu<0$, whereas the opposite holds for the branch $v_1 = -\mu$. 

\section{Concluding remarks}
\label{sec-con}

Because of their generality, we believe the results of Section \ref{sec-equiv},
involving the description of smooth planar curves as equivalence classes of submersions, 
to be of potential interest in different branches of applied mathematics. 
In the context of nonlinear circuit theory,
this formalism has led to 
a framework where homogeneous forms for incremental magnitudes can be handled systematically, making it 
easier to analyze different
properties of circuits with implicit characteristics, both in the
classical and in the memristive context. Such analyses
admit several extensions: let us mention, in particular, that 
all the results concerning
dynamic circuits can be easily extended to accommodate implicit characteristics
in reactive devices, and that most ideas seem to be applicable to circuits with
memcapacitors and meminductors.
Distributed systems can be possibly included in the same formalism. Our approach should be useful in other 
qualitative studies: in the scope of future research is the 
formulation of Routh-Hurwitz criteria for electrical circuits in graph-theoretic terms,
using the general multihomogeneous description of the characteristic polynomial here
discussed. This should be of help in a general analysis of stability problems in nonlinear circuits.
The study of other bifurcations, including bifurcations without parameters
in memristive circuits, may also benefit from our approach.

%Finally, we claim without proof than an extension of Theorem \ref{th-main} characterizes
%the nondegeneracy of DC operating points in circuits including also capacitors and inductors.
%Specifically, if the latter are %these reactive devices are assumed to be 
%governed by equations
%of the form $C(v_c)v_c' =i_c$, $L(i_l)i_l'=v_l$,
%with non-singular %capacitance and inductance matrices
%$C(v_c)$, $L(i_l)$, DC operating points are equilibria
%defined by the conditions $i_c=0$, $v_l=0$.
%% (together, of course, with Kirchhoff laws and the
%%characteristic equations of resistors). 
%Assuming again a fully-implicit form
%for resistors, and the absence
%%under the assumption that the circuit has neither 
%of loops defined by voltage sources
%and/or inductors and cutsets formed by current sources and capacitors,
%the geometric isolation of such operating points is guaranteed by the non-vanishing
%of the function %depicted 
%in (\ref{kir}), provided that the sum now ranges over 
%the circuit spanning trees which include all voltage sources and inductors but neither current sources
%nor capacitors. In passive contexts 
%and after dehomogenization this particularizes to an essentially 
%known result in circuit theory \cite{hagg}.
%Note that in this context
%and (ii) 
%the products still range only over resistive branches.

%\vspace{-0.1mm}

\end{document}